\documentclass[fleqn,usenatbib]{mnras}

\usepackage{newtxtext,newtxmath}

\usepackage[T1]{fontenc}

\DeclareRobustCommand{\VAN}[3]{#2}
\let\VANthebibliography\thebibliography
\def\thebibliography{\DeclareRobustCommand{\VAN}[3]{##3}\VANthebibliography}

\usepackage{graphicx}	
\usepackage{amsmath}	
\usepackage{longtable}
\usepackage{multirow}
\usepackage{cleveref}

\newcommand{\dsct}{\mbox{$\delta$~Scuti}}
\newcommand{\echelle}{{\'e}chelle}

\newcommand{\gdor}{\mbox{$\gamma$~Doradus}}

\newcommand{\muhz}{\mbox{$\mu$Hz}}
\newcommand{\Dnu}{\mbox{$\Delta\nu$}}

\newcommand{\numax}{\texorpdfstring{\mbox{$\nu_{\rm max}$}}{numax}}

\newcommand{\kepler}{{\em Kepler\/}}

\newcommand{\gaia}{{\em Gaia\/}}

\newcommand*{\myeqref}[2][equation~]{%
  \hyperref[{#2}]{#1(\ref*{#2})}%
}
\def\equationautorefname#1#2\null{%
  equation#1(#2\null)%
}

\newcommand{\autorefA}[1]{\hyperref[#1]{Appendix~\ref*{#1}}}
\newcommand{\orcidlink}[1]{\protect\href{https://orcid.org/#1}{\protect\includegraphics[width=8pt]{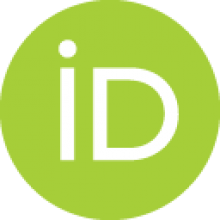}}}

\newcommand{\new}[1]{{\color{red}\textbf{#1}}} 
\renewcommand{\new}[1]{#1} 

\newcommand{\newb}[1]{{\color{red}\textbf{#1}}} 
\renewcommand{\newb}[1]{#1} 

\title[Halo Photometry for 98 Bright TESS Stars]{Halo Photometry and Asteroseismology for 98 of the Brightest Stars Observed by TESS}

\author[Rudrasingam et al.]{Jonatan Rudrasingam{\orcidlink{0009-0007-7973-9228}},$^{1}$\thanks{E-mail: jrud0912@uni.sydney.edu.au}
Timothy R. Bedding{\orcidlink{0000-0001-5222-4661}},$^{1}$
Benjamin J. S. Pope{\orcidlink{0000-0003-2595-9114}},$^{2}$
May Gade Pedersen{\orcidlink{0000-0002-7950-0061}},$^{1}$
\newauthor
Mikkel N. Lund{\orcidlink{0000-0001-9214-5642}},$^{3}$ 
Timothy R. White{\orcidlink{0000-0002-6980-3392}},$^{4}$ 
and Daniel Hey{\orcidlink{0000-0003-3244-5357}}$^{5}$
\\
$^{1}$Sydney Institute for Astronomy, School of Physics, University of Sydney, Sydney, NSW 2006, Australia\\
$^{2}$School of Mathematical and Physical Sciences, Macquarie University, 12 Wally's Walk, Macquarie Park, NSW 2113, Australia\\
$^{3}$Stellar Astrophysics Centre, Department of Physics and Astronomy, Aarhus University, Ny Munkegade 120, DK-8000 Aarhus C, Denmark \\
$^{4}$Sydney Informatics Hub, Core Research Facilities, University of Sydney, NSW 2006, Australia \\
$^{5}$Institute for Astronomy, University of Hawai`i, 2680 Woodlawn Drive, Honolulu, HI 96822, USA
}

\date{Accepted 2026 February 24. Received 2026 February 23; in original form 2026 January 22}

\pubyear{\the\year{}}

\begin{document}
\label{firstpage}
\pagerange{\pageref{firstpage}--\pageref{lastpage}}
\maketitle

\begin{abstract}
The Transiting Exoplanet Survey Satellite (TESS) mission has facilitated studies of asteroseismology, eclipsing binaries, and transits in many stars. However, the brightest stars saturate TESS, yet they are the most amenable to photon-hungry high-resolution studies and have long observational histories. In this work, we adapted the halo photometry used in $K$2 to extract light curves from the unsaturated halo pixels of the star's point spread function. We used this method to extract light curves for 98 of the brightest stars observed by TESS in Sectors 1--93. These bright stars include 15 red giants, five \dsct{} variables, eight stochastic low-frequency variables, eight eclipsing binaries, and 46 other variables. We measured \numax{} for 13 red giants using \textsc{pyMON} and \Dnu{} for one of them, $\beta$~Gem (Pollux). For five of them, this represents the first time that oscillations were detected. We derived their stellar masses using the measured \numax{} and previous interferometric and radiometric angular diameters. We also discovered \dsct{} and \gdor{} variability in $\alpha$~Cep, possible asteroseismic binary signatures in $\epsilon$~Car, and a new eclipsing binary, $\gamma$~And. Furthermore, we identified 18 stars in our sample that will be observed by the future PLAnetary Transits and Oscillations of stars (PLATO) mission, and 69 stars that have Stellar Observations Network Group (SONG) observations, including some simultaneous with TESS. The light curves are publicly available on the Mikulski Archive for Space Telescopes.









\end{abstract}




\section{Introduction}

The Transiting Exoplanet Survey Satellite (TESS), launched in 2018, has now observed nearly the entire sky, providing high-precision photometry for many bright, nearby stars \citep{Ricker2014, Ricker2015}. This has allowed many studies of oscillating stars, eclipsing binaries, and planetary transits \citep[see review by][]{Winn2024}. For some regions in the sky, TESS has now provided longer time series than \kepler{}. However, the TESS detector begins to saturate around seventh magnitude and electrons bleed into neighbouring pixels. Depending on the magnitude, the star can produce long bleed columns. Although studies have been made using \new{TESS} photometry for \new{solar-like oscillators} up to at least second magnitude \citep[e.g.][]{Lund2025}, \new{and O/B and A/F stars up to zeroth magnitude \citep[e.g.][]{Pedersen2019,Antoci2019,Bowman2022, Sharma2022,Rieutord2024}}, there are no systematic studies dealing with all the brightest stars \new{observed by TESS}. 

Bright stars are important because it is possible to do complementary observations including: interferometry \citep[e.g.][]{Baines2023}; spectroscopy \citep[e.g.][]{Strassmeier2018}; spectropolarimetry \citep[e.g.][]{Zwintz2020}; long-term radial-velocity (RV) monitoring of stellar companions, exoplanets, or unknown stellar variability \citep[e.g.][]{Hatzes2018}; and ground-based asteroseismic observations, either simultaneously \citep[e.g.][]{Kjeldsen2025} or otherwise \citep[e.g.][]{Arentoft2019, Knudstrup2023}. Furthermore, some of these stars have been observed by other space-based missions such as the $52$-mm star camera on Wide Field Infrared Explorer \citep[WIRE;][]{Buzasi2000a, Buzasi2000b}, the Solar Mass Ejection Imager on the Coriolis satellite \citep[SMEI;][]{Jackson2004}, the Microvariability and Oscillation of Stars \citep[MOST;][]{Walker2003}, and the BRIght Target Explorer (BRITE)-Constellation \citep{Weiss2014, Pablo2016}. Obtaining TESS photometry can help validate and improve previous results from these missions.

For \kepler{} and $K$2, different methods have been developed to deal with saturated stars. One of these, smear photometry \citep{Pope2016, Pope2019a}, uses collateral smear measurements to obtain the light curves for these stars, with the downside that it cannot distinguish stars in crowded fields. It has been used to extract the light curves of 102 saturated stars observed by \kepler{} \citep{Pope2019a}. The other, more precise, method is halo photometry \citep{White2017, Pope2019}. Halo photometry works by performing photometry in the halo of the point spread function (PSF), where weights are fitted to each pixel by minimizing an objective function. Halo photometry was inspired by the approach used by \citet{Aerts2017} for the O9.5Iab star HD 188209 and has similarities to the Optimized-Weight Linear (OWL) photometry concept described by \citet{Hogg2014}. Halo photometry has successfully been used with $K$2 for: the seven brightest stars in the Pleiades \citep{White2017}, $\alpha$~Tau \citep[Aldebaran;][]{Farr2018}, $\rho$~Leo \citep{Aerts2018}, $\iota$~Lib \citep{Buysschaert2018}, $\epsilon$~Tau \citep{Arentoft2019}, and 162 saturated stars observed in $K$2 \citep{Pope2019}. With TESS it has been used for $\alpha$~Aql \citep[Altair;][]{Rieutord2024}, and the two brightest stars in the Scorpius-Centaurus Association, $\beta$~Cen and $\alpha$~Cru \citep{Sharma2022}. However, halo photometry can be limited by both crowding, which dilutes the stellar signal in the halo pixel, and variations in the background, as shown for the third-magnitude star $\tau$~Ceti \citep{Eisner2019}, which caused a false transit to appear in the halo photometry light curve.

In this paper, we perform halo photometry on 98 of the brightest saturated stars observed by TESS in Sectors 1--93. These stars range from the brightest second-magnitude stars up to most of the zeroth-magnitude stars. They include almost all of the brightest red giants (RGs), where it is possible to determine \numax{} for 13 of them, allowing us to estimate masses. The sample also includes some of the brightest known \dsct{} stars, \gdor{} stars, slowly pulsating B-type (SPB) stars, $\beta$~Cephei variables, stochastic low-frequency variables (SLFs), eclipsing binaries (EBs), as well as other types of variable stars such as long-period variables (LPVs) and $\alpha^2$ CVn variables. For some of these stars, this study reports the first detection of stellar variability.

Among the stars in our sample, we found three highlights. The first is the second magnitude A7IV-V star $\alpha$~Cep (Alderamin), which we discovered to be a $\delta$~Sct-$\gamma$~Dor hybrid pulsator using 11 TESS sectors. Secondly, we found that the binary system $\epsilon$~Car (Avior) is either an asteroseismic binary consisting of an LPV and an SPB star,  or an LPV star with an $\alpha^2$ CVn companion. Lastly, there is the multiple system $\gamma$~And (Almach), in which we discovered two of the components to be eclipsing.

We structure this paper as follows. In \autoref{sec:sample}, we describe the selection process of our sample. In \autoref{sec:lcurves}, we describe the extraction of our light curves using halo photometry and comparison with pipeline light curves. We will then present results for oscillating RGs (\autoref{sec:res_rg}), pulsating A/F stars (\autoref{sec:res_af}), pulsating O/B stars (\autoref{sec:res_ob}), and EBs (\autoref{sec:res_eb}). In \autoref{sec:discussion}, we compare the halo photometry results from TESS with the $K$2 Bright Star Survey \citep{Pope2016} and compare the halo photometry amplitudes with previous measurements for EBs. In \autoref{sec:future_work}, we explore potential synergies with the PLAnetary Transits and Oscillations of stars \citep[PLATO;][]{Rauer2025} and the Stellar Observations Network Group \citep[SONG;][]{Andersen2014, Grundahl2017, Kjeldsen2025}.

\section{The TESS Halo Sample}\label{sec:sample}

We selected our sample using the Hipparcos and Tycho catalogues \citep{Perryman1997,Hoeg1997,Leeuwen1997,hip1997}, since the \gaia{} catalogue \citep{Gaia2023} is not complete for bright stars. We selected all stars with magnitudes brighter than 2.66 in $V$. We chose this cut to avoid overlap with the ``TESS Luminaries Sample'' of solar-like-oscillators by \citet{Lund2025}, whose brightest star is the subgiant $\eta$ Boo ($V$ = 2.68). Since they successfully used a standard photometric method to extract the light curves for their stars, we did not see the need to include fainter stars.


We excluded the eight brightest stars ($V \le 0.40$) because they require special treatment, due to the poor smear correction and a lack of halo pixels containing the stellar variability (Sirius, Canopus, $\alpha$~Cen, Arcturus, Vega, Capella, Rigel, and Procyon). We also excluded two stars, $\alpha$~Peg (Markab) and $\beta$~Sco (Acrab), that were not observed by TESS as of Sector 93. Finally, we added the RG $\beta$~Her (Kornephoros) to our sample, despite it having an apparent magnitude of $V=2.78$, because it had a small target pixel file (TPF), which did not encompass the entire bleed. We include a sample of 98 stars in this paper, with data from a total of 411 TESS sectors.



\begin{figure}
 \includegraphics[width=\columnwidth]{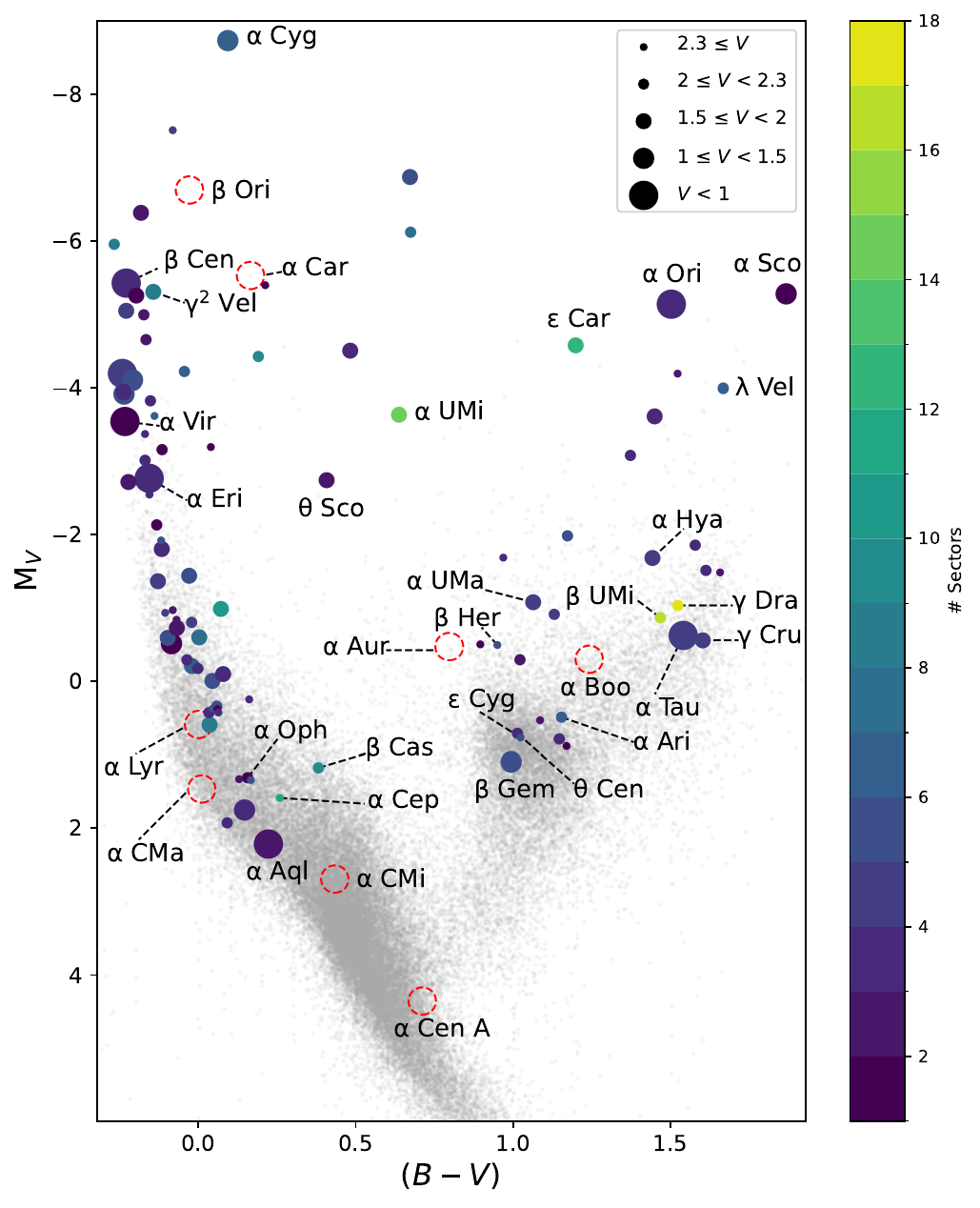}
 \caption{A CMD showing the stars in our sample in the Hipparcos and Tycho catalogues, coloured based on the number of TESS sectors available, overlaid on the background of the Hipparcos stars. The size of the dots corresponds to the $V$ \new{mag}. The red circles are the brightest stars, which we excluded from our sample. \new{Various} prominent and interesting stars are highlighted in the figure.}
 \label{fig:cmd}
\end{figure}

Using both the Hipparcos parallaxes and apparent $V$-band magnitude, we calculated the absolute magnitude ($M_V$). We used the calculated $M_V$ and the colour $(B - V)$ to plot the stars in a colour-magnitude diagram (CMD) shown in \autoref{fig:cmd}, where we colour-code the stars by the number of TESS sectors available. The CMD shows that most of our stars are either O-B-A-F dwarfs or evolved stars such as RGs. We also see from \autoref{fig:cmd} that the number of available TESS sectors ranges from 1 to 18; none of the stars in our sample are in the northern or southern TESS continuous viewing zone (CVZ). We list the stars in \autoref{tab:stars_list}, providing their names, TIC, HD, $V$, $B - V$, $M_V$, spectral types, number of sectors ($N$), number of good \textsc{halophot} sectors ($N_H$, see \autoref{sec:discussion_quality}), number of good pipeline sectors ($N_P$, see \autoref{sec:discussion_quality}), and variability classification. We obtained the spectral types from the Yale Bright Star Catalog \citep[5th ed.;][]{Hoffleit1995}.

\section{Processing of the light curves}\label{sec:lcurves}

\subsection{Halo photometry}\label{sec:halophotometry}

Halo photometry measures flux variations in the unsaturated halo of the PSF by masking the saturated pixels. The remaining pixels are then assigned weights by minimising an objective function, whereafter a light curve is created by summing the weighted pixels \citep{White2017, Pope2019}. Halo photometry has the advantage over standard aperture photometry by ignoring the saturated pixels and not needing larger aperture mask to encompass the bleed. 
We provide a brief summary of the method. 

We can describe the halo flux in a given image as:
\begin{equation}
f_i = \sum^N_{j=1}w_jp_{ij},
\end{equation}
where $i$ is the image index, $N$ is the number of pixels, $w_j$ is the weight of the pixel $j$, and $p_{ij}$ is the flux of pixel $j$ in the image $i$. All of the weights are defined to be non-negative and must sum to unity. The weights are assigned to the pixels by minimising an objective function, chosen to be Total Variation (TV) given by:
\begin{equation}
TV = \frac{\sum^M_{i=1}|f_i - f_{i-1}|}{\sum^M_{i=1}f_i}, 
\end{equation}
where $M$ is the total number of observations. See \citet{Pope2019} for more details.

For this work, we used an implementation of halo photometry closely following the \textsc{Python} module \textsc{halophot}\footnote{\url{https://github.com/hvidy/halophot}}, with a modification in solving the objective function compared to \cite{Pope2019}. We used the \textsc{Python} framework \textsc{Jax}\footnote{\url{https://github.com/jax-ml/jax}} \citep{jax2018} to both calculate and solve the objective optimisation instead of \textsc{autograd} and \textsc{scipy}, as was done by \cite{Pope2019}. The objective function itself was solved by using the L-BFGS algorithm implemented in the \textsc{JAX} module \textsc{JAXopt}\footnote{\url{https://github.com/google/jaxopt}} \citep{jaxopt_implicit_diff}. We also note that our pixel masking differs slightly from \textsc{halophot}, in that we also exclude the pixels immediately above and below the saturated pixels. We did this to avoid the pixels at the edge of the bleed columns, which would otherwise inflate the amplitude of the stellar variability. All of these changes have been implemented in the current version of \textsc{halophot}.

\subsection{Extraction of the light curves}\label{sec:extract_lcurves}

To obtain the TPFs of all of the stars in our sample, we used TESScut \citep{Brasseur2019} with a pixelcut of $31\times31$ in the Full Frame Images (FFIs). We chose this dimension to obtain a sufficient number of halo pixels and include the background around the star, noting there is no need to encompass the entire bleed column. 

To fit the background model, we used the \textsc{Python} module \textsc{tessbkgd}\footnote{\url{https://github.com/hvidy/tessbkgd}}, which fits a spatially varying model using the background pixels. It has previously been used to fit the background of the third magnitude star $\iota$~Dra \citep{Hill2021, Campante2023}. Unlike those studies, we fitted the background with a linear model instead of a second-order polynomial, but we note that the choice of model had minimal effect on the extracted light curves. We then used the background fit from \textsc{tessbkgd} to remove all pixels for which the ratio between the fitted background and the median flux was greater than 0.3. We did this to remove most of the background stars, which could cause contamination. For the bright background stars, we masked them after inspecting the TPFs visually. 

For the halo photometry, we ran \textsc{halophot} twice. The first fit was used to identify and remove outlier cadences and those with zero flux via sigma clipping. We also removed sections affected by earthshine and moonshine. \new{These cadences have high flux in the beginning and end of each TESS orbit, and were removed by manually masking these cadences after visual inspecting the initial light curves.} We did this to improve our \textsc{halophot} fit, since these epochs would dominate the light curve variability and influence the TV fit. A few of the TPFs contained single pixels with high weights, which we masked to prevent them from dominating the light curve. \new{We then recalculated the pixel weights for the second and final \textsc{halophot} fit.}

\subsection{Quality of halo photometry}\label{sec:discussion_quality}

To evaluate the quality of the extracted \textsc{halophot} light curves, we downloaded the pipeline light curves from Mikulski Archive for Space Telescopes (MAST)\footnote{\url{https://mast.stsci.edu/portal/Mashup/Clients/Mast/Portal.html}} using the \textsc{Python} module \textsc{lightkurve} \citep{lightkurve2018}. We used the 120-s cadence light curves from TESS Science Processing Operations Center \citep[SPOC;][]{Jenkins2016}, when they were available, and TESS-SPOC light curves from the FFIs \citep{Caldwell2020} when they were not. If the TESS-SPOC light curve was also not available, we used light curves from the Quick-Look Pipeline \citep[QLP;][]{Huang2020a, Huang2020b, Kunimoto2021, Kunimoto2022}. In the few cases where the QLP light curve was also unavailable (6 sectors), we created our own custom aperture photometry using \textsc{lightkurve}, using the same available pixels as in \textsc{halophot} but without TV minimisation.

We used the method described in Section~4 of \citet{Kjeldsen2025} to de-trend the time series using a high-pass filter and remove outliers. We tailored the high-pass filter width for each star to preserve the stellar signal in the light curves. Afterwards, we assigned variability classification to all of our stars by looking at their light curves and Fourier spectra from both \textsc{halophot} and pipeline. From the \textsc{halophot}, SPOC, QLP, and custom aperture light curves, we found that 77 out of 98 stars show variability.

\begin{figure*}
 \includegraphics[width=1.5\columnwidth]{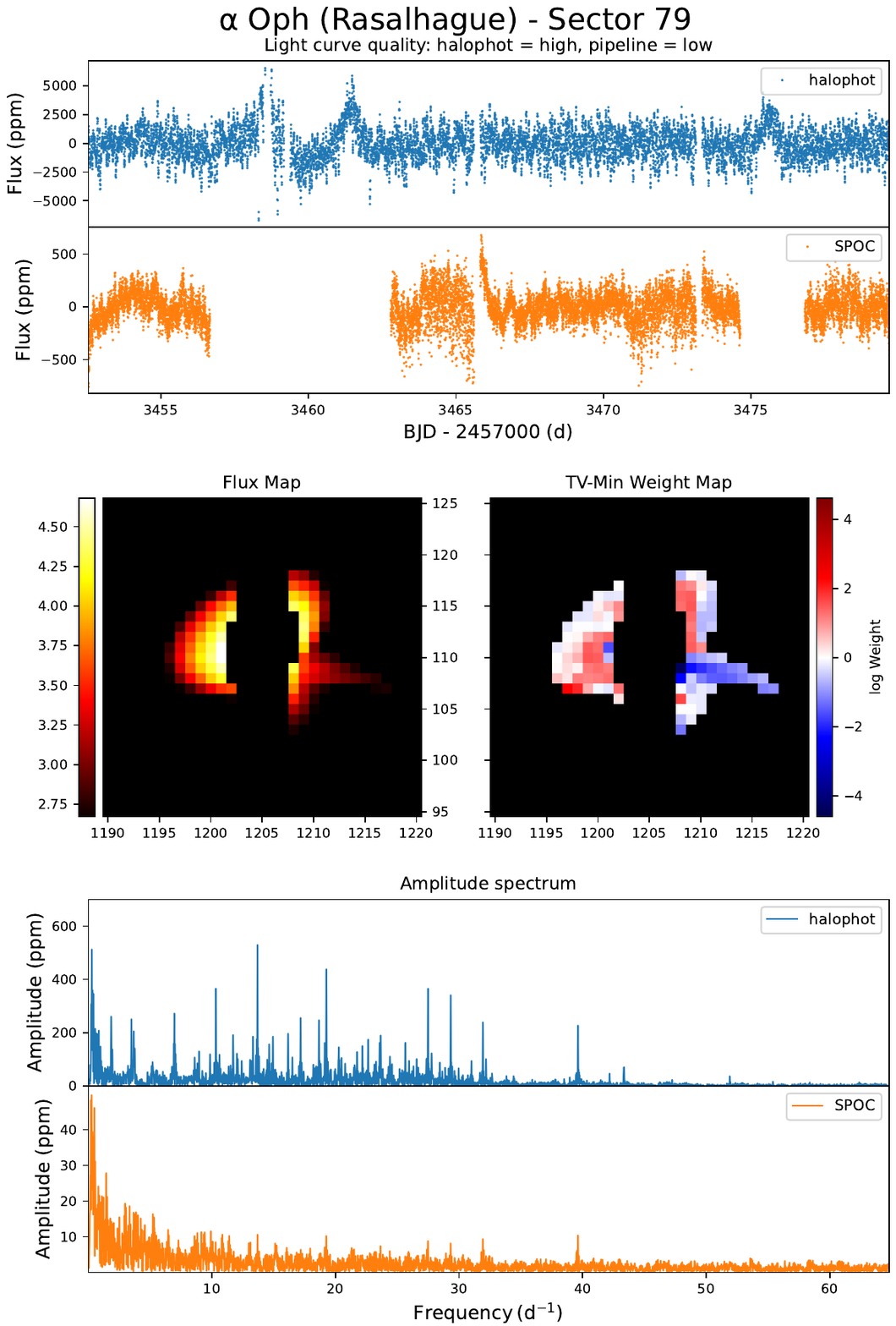}
 \caption[]{Single-sector diagnostic plots of the star $\alpha$~Oph (Rasalhague) in Sector 79. The top panel shows the \textsc{halophot} light curve in blue. The second panel shows the light curve from SPOC in orange. The middle two panels show the log-flux map (left) after masking of the saturated pixels and weightmap (right). The two bottom panels show the amplitude spectrum from both the \textsc{halophot} light curve and the SPOC light curve. The quality of the \textsc{halophot} and pipeline light curves are shown below the title.}
 \label{fig:rasalhague}
\end{figure*}

\begin{figure*}
 \includegraphics[width=1.5\columnwidth]{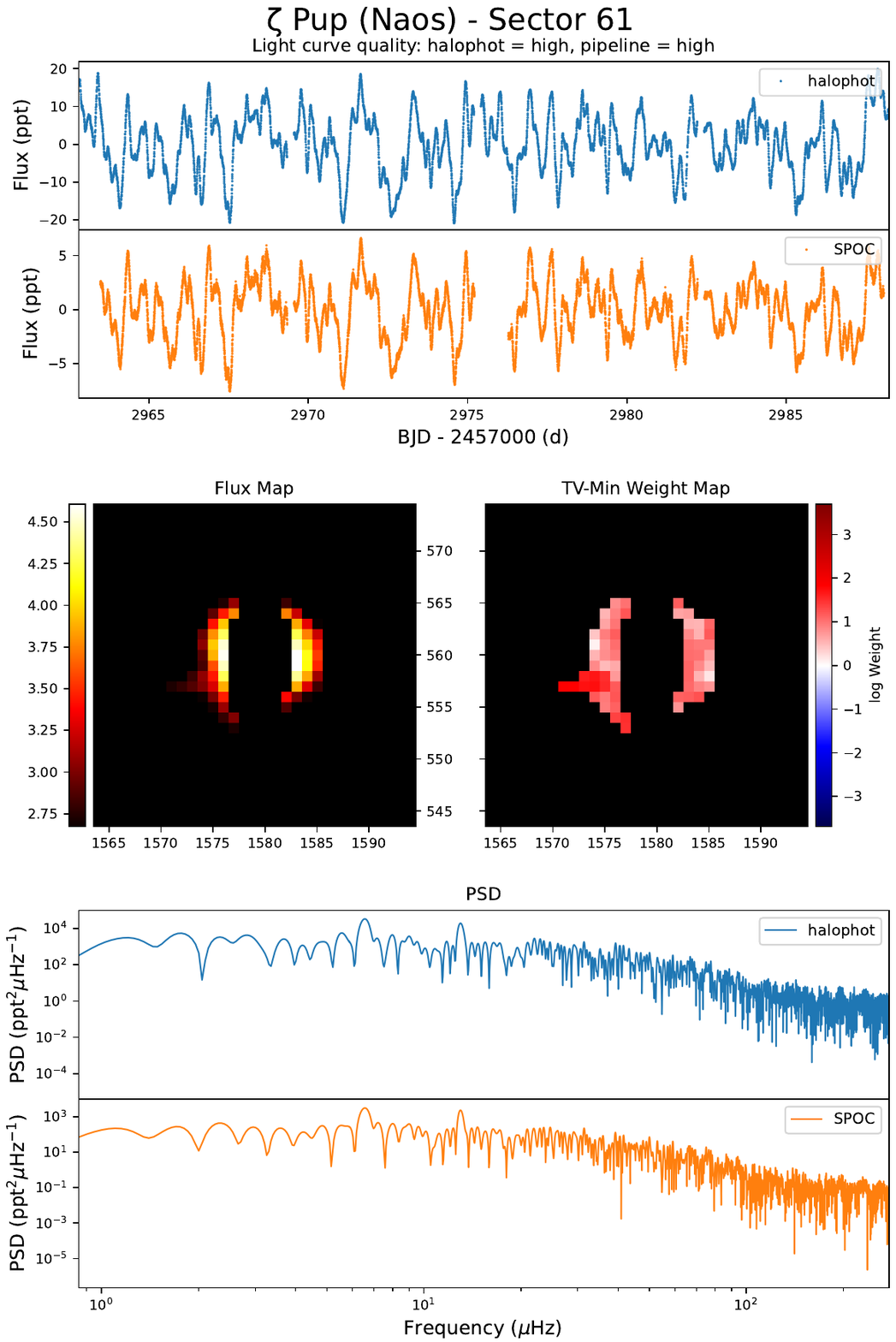}
 \caption{Similar single-sector diagnostic plots as \autoref{fig:rasalhague}, but for the SLF $\zeta$~Pup (Naos) in Sector 61. The only difference is that we plot the power spectral density on a log scale rather than the linear amplitude spectra in the bottom panels.}
 \label{fig:naos}
\end{figure*}

\begin{figure*}
 \includegraphics[width=1.5\columnwidth]{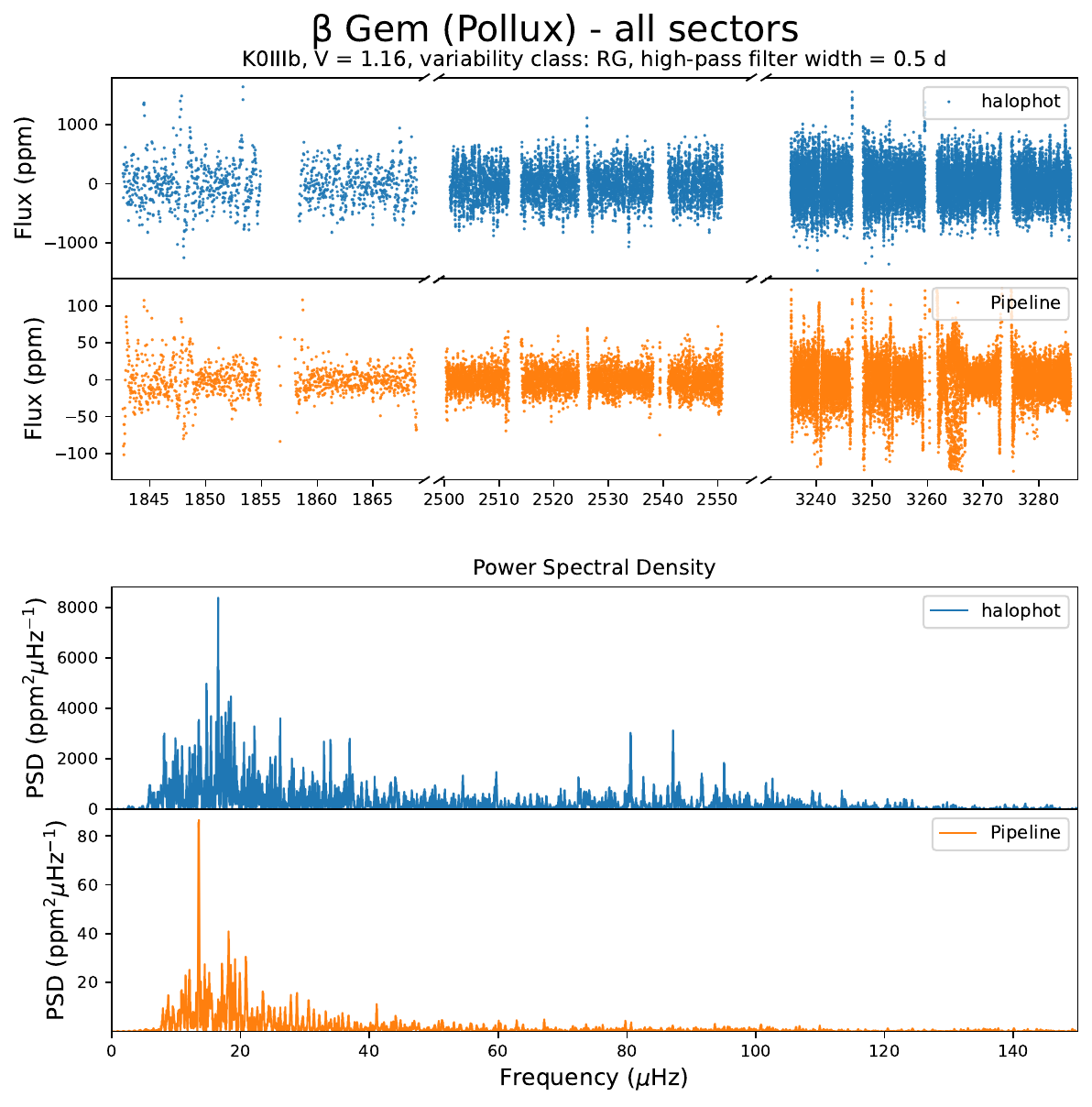}
 \caption{All-sector diagnostic plots of the RG $\beta$~Gem (Pollux). The top panel shows the \textsc{halophot} light curve in blue. The second panel shows the light curve from the pipeline in orange. In this case the pipeline light curves are exclusively from QLP. The two bottom panels show the power spectral density from both the \textsc{halophot} light curve and the pipeline aperture light curve. The spectral type, $V$ magnitude, variability class, and the width of the high-pass filter are shown below the title.}
 \label{fig:corhydræ}
\end{figure*}

We made two kinds of diagnostic plots for our targets: single-sector and all-sectors. We show examples of single-sector diagnostic plots in \autoref{fig:rasalhague} for the $\delta$~Sct variable $\alpha$~Oph (Rasalhague) in Sector 79, and in \autoref{fig:naos} for the SLF variable $\zeta$~Pup (Naos) in Sector 61. In \autoref{fig:corhydræ} we show an example of an all-sectors plot for the RG $\beta$~Gem (Pollux) across all available TESS sectors. Each diagnostic plot shows the time series and Fourier spectrum from both the \textsc{halophot} and pipeline light curves. The all-sectors plots contain the spectral type, the $V$ magnitude, our variability classification, and the width of the high-pass filter used for de-trending. The single-sector diagnostic plots contain the time series, the flux map, the \textsc{halophot} weightmap, and the power spectra, for the given sector. They also contain information about the quality of \textsc{halophot} and pipeline light curves. We compiled three sets of diagnostic plots: the first for classical pulsators and EBs, the second RGs and LPVs, and the third for SLFs. All the diagnostic plots are available on GitHub\footnote{\url{https://github.com/JonatanRudrasingam/TESS_halophot}}.

To compare the \textsc{halophot} light curves with SPOC, TESS-SPOC, QLP, and with our custom non-TV-minimised light curves, we examined both the time series and the power spectrum to assess the quality. More specifically, we examined the scatter and systematics in the time series, the signals in the power spectrum and, for multi-sector stars, the quality of the light curve and Fourier spectrum relative to other sectors. For each sector, we rated the \textsc{halophot} and pipeline light curves as either good or poor. The number of good \textsc{halophot} light curves, $N_H$, and the number of good pipeline light curves, $N_P$, are listed in \autoref{tab:stars_list} for each star.

Our inspections of the results revealed several trends. Firstly, the QLP light curves for our sample are nearly always of lower quality than the \textsc{halophot} light curves. This can be explained by the circular aperture mask used by QLP, which is not well-suited to saturated stars. For stars without SPOC or TESS-SPOC light curves, the best available light curves are therefore from \textsc{halophot}. Secondly, the best SPOC light curves are comparable to the best \textsc{halophot} light curves, although about 79 SPOC light curves have poorer quality than the \textsc{halophot} counterpart, especially at the bright end of our sample. SPOC light curves can be degraded by either poor aperture masks or TPFs that fail to encompass the entire bleed column. For very bright stars, the saturated columns will not only be large, requiring larger TPFs and aperture, but they can also show non-linear effects, complicating the photometry. An exception is the brightest star in our sample, $\alpha$~Eri (Achernar), which has a single SPOC light curve, that is comparable to \textsc{halophot}. Meanwhile, 35 \textsc{halophot} light curves are worse than their SPOC counterparts, which can be explained by contamination from background stars. For example, this is the case for \new{the recently discovered $\alpha^2$~CVn variable} $\beta$~Tau \new{\citep{Begari2026}, where multiple background stars contaminate the unsaturated halo pixels.} 

For some variable stars, the SPOC light curves have lower amplitudes than the \textsc{halophot} even when both light curves appear to have comparable quality. The low amplitudes occur because the SPOC aperture mask does not fully encapsulate the bleed columns, which dilutes the stellar variations. Examples of this can be seen in \autoref{fig:rasalhague} for $\alpha$~Oph, and in \autoref{fig:naos} for $\zeta$~Pup, where the SPOC amplitudes are significantly lower. For stars that have comparable light curves, like $\zeta$~Pup in Sector 61, but different amplitudes, we rated the light curves equally. However, we recommend using \textsc{halophot} light curves, especially if the amplitude is important for the given study. Whether the amplitudes from \textsc{halophot} are reliable is discussed in \autoref{sec:discussion_amplitude}.


For the $\beta$~Cep star $\alpha$~Cru, we assigned poor quality to all the SPOC light curves, despite the light curves themselves being comparable to \textsc{halophot}. We made this choice because the SPOC amplitude spectrum has a peak at $\sim$0.81\,d$^{-1}$, which we attribute to the star HD 108250 ($V=4.8$), which is inside the bleed column of $\alpha$~Cru. Masking pixels close to the position of HD 108250 removed the signal from the \textsc{halophot} light curves. This is an interesting case in which contamination has a greater effect on aperture photometry than on halo photometry.

For most of our EBs, except for $\alpha$~Gem (Castor) and $\lambda$~Sco, the \textsc{halophot} light curves are generally better than the pipeline light curves, with larger eclipse depths that are consistent with published values (see \autoref{sec:discussion_amplitude}). For the RGs and LPVs, our \textsc{halophot} light curves are generally better than pipeline light curves because most of them only have QLP light curves. Another detail is that stars fainter than $V=2.55$ mostly either have comparable or worse \textsc{halophot} light curves than SPOC, with a few exceptions. This result is consistent with the fact that the aperture photometry method works for stars $\geq$ 2.68\,$V$ \citep{Lund2025}. An exception is the RG $\beta$~Her ($V = 2.78$), where all 5 SPOC light curves have poorer quality than the \textsc{halophot} light curves due to small TPFs (see \citealt{Lund2025}). So \textsc{halophot} light curves are almost always better than QLP light curves, and they are comparable to SPOC light curves, with \textsc{halophot} being better for the brightest stars, and SPOC being better for the fainter stars. However, the amplitudes for SPOC are sometimes lower than \textsc{halophot} due to poor aperture masks.


\section{Results for oscillating red giants}\label{sec:res_rg}

We identified 15 RGs in our sample based on their spectral type, position in the CMD, light curves, and power spectra. RGs are known to have stochastic oscillations excited and damped by the turbulence in the convective envelope, which is the same mechanism that drives oscillations in the Sun \citep{Chaplin2013}. These solar-like oscillators show characteristic features in their power spectra, which can be used to make inferences about their physical properties. The first is the large frequency separation, \Dnu{}, between modes with the same angular degree, while the second is the frequency at maximum power, \numax{}. \citet{Ulrich1986} showed that \Dnu{} scales approximately with the mean density of a star:
\begin{equation}
\Dnu{} \propto \sqrt{\tilde{\rho}}.
\end{equation}
Meanwhile, \numax{} scales approximately with the acoustic cut-off frequency of the atmosphere ($\nu_{\rm ac}$), which in turn is related to the surface gravity and the effective temperature \citep{Brown1991, Kjeldsen1995a, Belkacem2011}:
\begin{equation}
\numax{} \propto \nu_{\rm ac} \propto g/\sqrt{T_{\rm eff}}.
\end{equation}
Rearranging these two equations yields scaling relations for both mass and radius \citep{Stello2008, Kallinger2010}.
However, we were only able to determine \Dnu{} for one star, $\beta$~Gem (Pollux; see \autoref{sec:ind_rg_stars_pollux}). The remaining RGs have low \numax{} and too few sectors to resolve \Dnu{}. If only \numax{} is available, we can instead use a different scaling relation assuming the radius is known:
\begin{equation}\label{eq:scaling_m_numax}
\frac{M}{M_{\odot}} \approx \left(\frac{\nu_{\textrm{max}}}{\nu_{\textrm{max}, \odot}}\right)\left(\frac{R}{R_{\odot}}\right)^{2}\left(\frac{T_\textrm{eff}}{T_{\textrm{eff}, \odot}}\right)^{1/2}.
\end{equation}
Due to their brightness, all our RGs have published angular diameters and parallaxes, so \autoref{eq:scaling_m_numax} can be used. Using this scaling relation, we aimed to determine the approximate masses of the RGs in our sample without relying on detailed stellar modelling. 

We used published radii from interferometry, mostly obtained with the Navy Precision Optical Interferometer \citep[NPOI;][]{Armstrong1998}. For three stars that did not have NPOI measurements, $\alpha$~Phe, $\epsilon$~Sco, and $\theta$~Cen, we only had so-called radiometric angular diameters from \citet{Cohen1999}, which we used along with the Hipparcos parallax to derive their radii. For $T_{\rm eff}$, we used the same interferometric sources \citep{Baines2018, Baines2021, Baines2023, Baines2025}, except for Aldebaran \newb{\citep{Strassmeier2018}}, $\alpha$~Phe \citep{Charbonnel2020}, $\epsilon$~Sco \citep{Paegert2022}, and $\theta$~Cen \citep{Ottoni2022}\newb{. Note that some of the uncertainties are unrealistically small, but we kept with the published values since they make a negligible contribution to the error budget.}

To derive \numax{} from the TESS light curves, we first calculated the weighted power spectrum density using the Lomb-Scargle method \citep{Lomb1976, Scargle1982}, with weights computed as described in Section~4 of \citet{Kjeldsen2025}. We calculated the weights for each sector separately before combining the time series. To measure \numax{} we used the \textsc{Python} module \textsc{pyMON}\footnote{\url{https://github.com/maddyhowell/pyMON}} \citep{Howell2025}, which is a simplified version of the \textsc{SYD} pipeline \citep{Huber2009, Chontos2022}. We used an initial value of \numax{} determined by visual inspection of the power spectrum, and used a linear line in log-space as our background to derive \numax{} in \textsc{pyMON}. 

To estimate approximate uncertainties for \numax{}, we used Fig. 5 from \citet{Hon2021}, which shows the distribution of estimated fractional uncertainties for \numax{} from synthetic TESS data. Using this, we adopted uncertainties of 7 percent for \numax{} below 10\,\muhz{}, 6 percent for \numax{} between 10 and 20\,\muhz{}, 5 percent for \numax{} above 20\,\muhz{}, and 4 percent for $\beta$~Gem (Pollux), with its \numax{} of 89.25\,\muhz{}. 

\begin{figure*}
 \includegraphics[width=1.75\columnwidth]{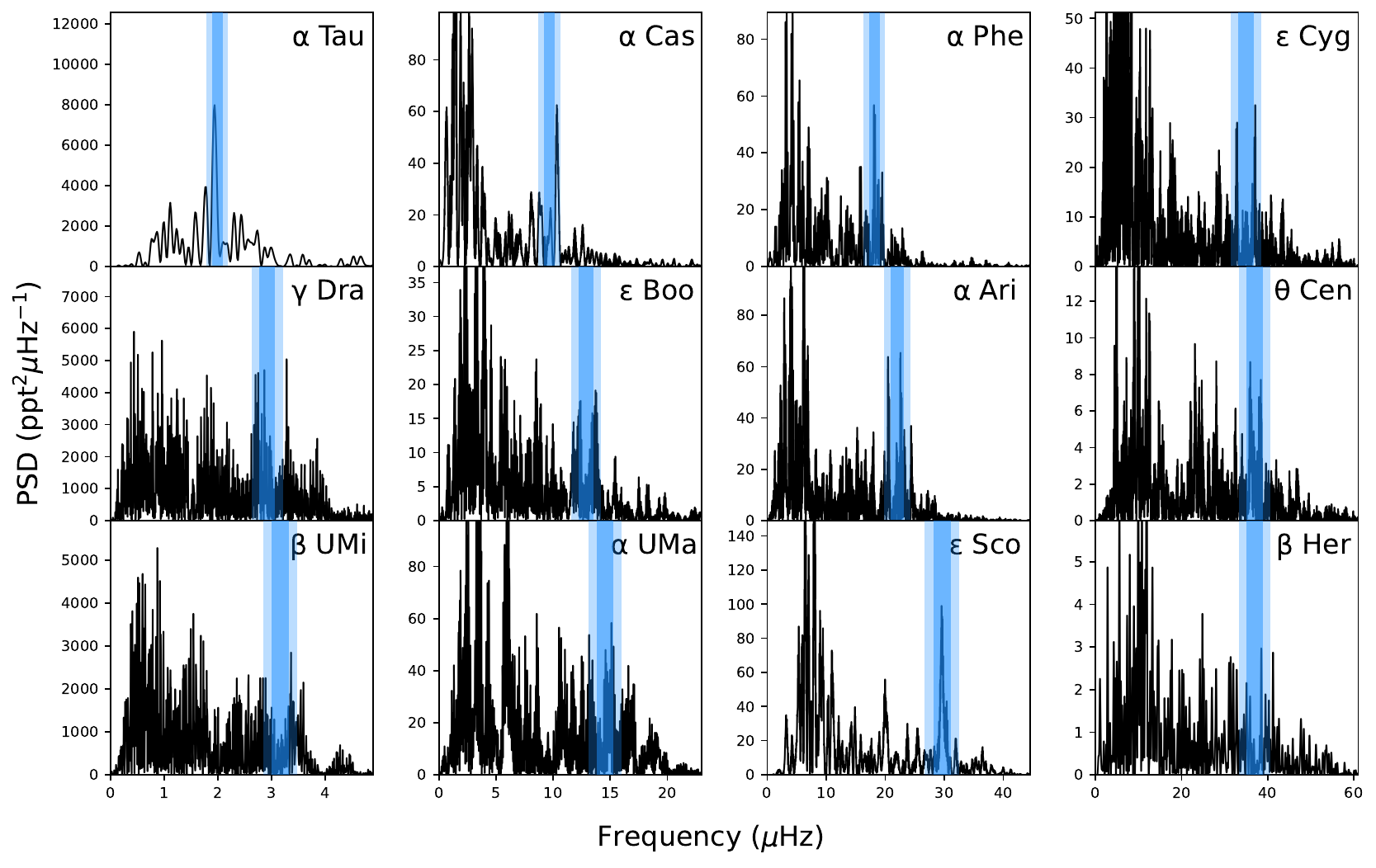}
 \caption{Power spectral density for 12 RGs with measurable \numax{}. The first column shows the PSD in the range 0--5\,\muhz{}, the second in the region 0--23\,\muhz{}, the third at 0--44.5\,\muhz{}, and the fourth at 0--61\,\muhz{}. The blue shaded regions indicate 1$\sigma$ and 2$\sigma$ uncertainties of the derived \numax{}. We note that $\beta$~Gem (Pollux) is not included in this figure.}
 \label{fig:astero_gallery}
\end{figure*}

\begin{figure}
 \includegraphics[width=\columnwidth]{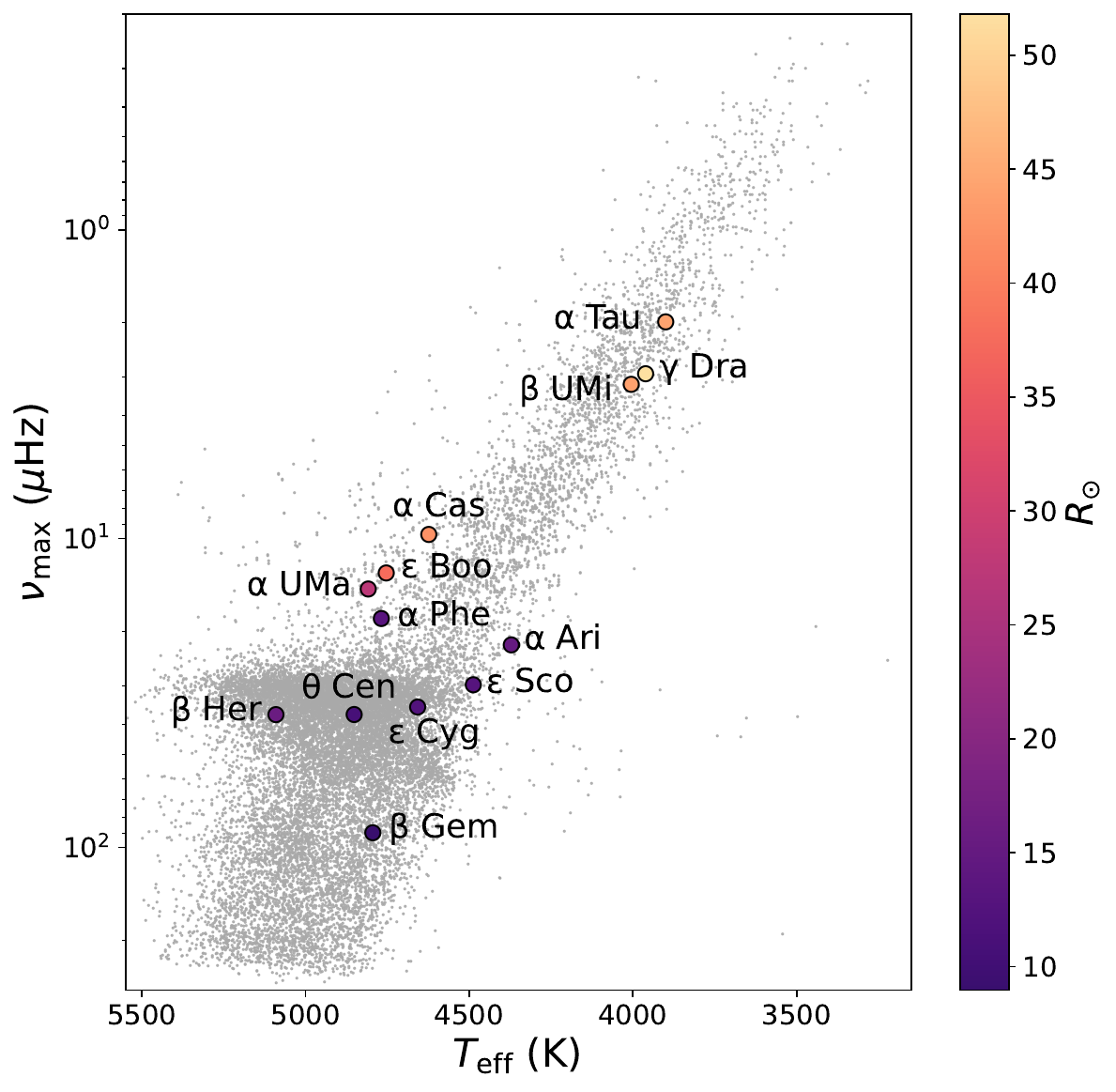}
 \caption{An asteroseismic HR diagram showing the \numax{} and $T_{\rm eff}$ of RGs in our sample colour-coded by their radius. Shown in grey are the RGs from \citet{Yu2018a} and the LPVs from \citet{Yu2020}.}
 \label{fig:HR_astero}
\end{figure}

We were able to determine \numax{} for 13 RGs. In \autoref{tab:gap} we list the RGs with measured \numax{} values, along with their $T_{\rm eff}$, $R$, $M$, the angular diameter references, and the reference for the first asteroseismic detection. Among these RGs, five are newly detected solar-like oscillators. In \autoref{fig:astero_gallery}, we show power spectra for 12 of them, along with their measured \numax{} values. Note that $\beta$~Gem (Pollux), which has a high \numax{} than the rest, is discussed separately in \autoref{sec:ind_rg_stars_pollux}. In \autoref{fig:HR_astero} we show the asteroseismic HR-diagram for the RGs colour-coded by their radius. 

For the remaining two RGs in our sample, $\gamma$~Leo (Algieba) and $\alpha$~Ser (Unukalhai), we were not able to measure \numax{}, despite detecting variability.
\begin{table*}
	\centering
	\caption{List of 13 RGs with \numax{} determined from \textsc{pyMON}.}
	\label{tab:gap}
	\begin{tabular}{lcccccc}
		\hline
		Star & \numax{} (\muhz{}) & $T_{\rm eff} (K) $ & $R \ (R_{\odot})$ & $M \ (M_{\odot})$ & Angular diameter reference & First asteroseismic discovery \\
		\hline
		$\alpha$~Tau & 1.99 $\pm$ 0.10 & 3900 \newb{$\pm$ 50}$^1$ & 44.20 $\pm$ 0.90 & 1.03  $\pm$ 0.07 & \citet{Farr2018} & \citet{Farr2018} \\ 
        $\gamma$~Dra & 2.92 $\pm$ 0.15 & 3964 \newb{$\pm$ 29} & 51.80 $\pm$ 0.26 & 2.10 $\pm$ 0.11 & \citet{Baines2021} & \citet{Hatzes1998a} \\
        $\beta$~UMi & 3.16 $\pm$ 0.16 & 4008 \newb{$\pm$ 37} & 44.13 $\pm$ 0.22 & 1.66 $\pm$ 0.09 & \citet{Baines2021} & \citet{Tarrant2008} \\
        $\alpha$~Cas & 9.68 $\pm$ 0.48 & 4625 \newb{$\pm$ 37} & 42.15 $\pm$ 1.55 & 4.98 $\pm$ 0.45 & \citet{Baines2025} & This work \\ 
        $\epsilon$~Boo & 12.90 $\pm$ 0.65 & 4755 \newb{$\pm$ 66} & 37.61 $\pm$ 1.38 & 5.36 $\pm$ 0.48 & \citet{Baines2021} & This work \\ 
        $\alpha$~UMa & 14.54 $\pm$ 0.73 & 4810 \newb{$\pm$ 22} & 27.33 $\pm$ 0.49 & 3.21 $\pm$ 0.20 & \citet{Baines2025} & \citet{Buzasi2000b} \\ 
        $\alpha$~Phe & 18.12 $\pm$ 0.91 & 4770 \newb{$\pm$ 250}$^2$ & 13.39 $\pm$ 0.29 & 0.96 $\pm$ 0.0\newb{7} & \citet{Cohen1999} & This work \\ 
        $\alpha$~Ari & 22.06 $\pm$ 1.10 & 4373 \newb{$\pm$ 12} & 15.19 $\pm$ 0.10 & 1.43 $\pm$ 0.08 & \citet{Baines2023} & \citet{Kim2006} \\
        $\epsilon$~Sco & 29.60 $\pm$ 1.50 & 4489 \newb{$\pm$ 40}$^\newb{3}$ & 12.91 $\pm$ 0.25 & 1.41 $\pm$ 0.09 & \citet{Cohen1999} & \citet{Kallinger2019} \\ 
        $\epsilon$~Cyg & 34.94 $\pm$ 1.75 & 4659 \newb{$\pm$ 6} & 12.41 $\pm$ 0.29 & 1.56 $\pm$ 0.11 & \citet{Baines2023} & \citet{Kallinger2019} \\ 
        $\theta$ Cen & 36.95 $\pm$ 1.85 & 4853 \newb{$\pm$ 41}$^\newb{4}$ & 10.96 $\pm$ 0.20 & 1.32 $\pm$ 0.08 & \citet{Cohen1999} & This work \\ 
        $\beta$~Her & 36.98 $\pm$ 1.85 & 5092 \newb{$\pm$ 64} & 15.92 $\pm$ 0.41 & 2.85 $\pm$ 0.21 & \citet{Baines2018} & This work \\
        $\beta$~Gem & 89.25 $\pm$ 3.57 & 4796.0 \newb{$\pm$ 10} & 8.92 $\pm$ 0.3\newb{7} & 2.09 $\pm$ 0.2\newb{6} & \citet{Baines2025} &  \citet{Hatzes2007}\\ 
		\hline
        \multicolumn{7}{l}{1: \newb{\citet{Strassmeier2018}}, 2: \citet{Charbonnel2020}, \newb{3: \citet{Paegert2022}, 4: \citet{Ottoni2022}}}
	\end{tabular}
\end{table*}


\subsection{$\alpha$~Tau (Aldebaran)} \label{sec:ind_rg_stars_aldebaran}

\begin{figure}
 \includegraphics[width=\linewidth]{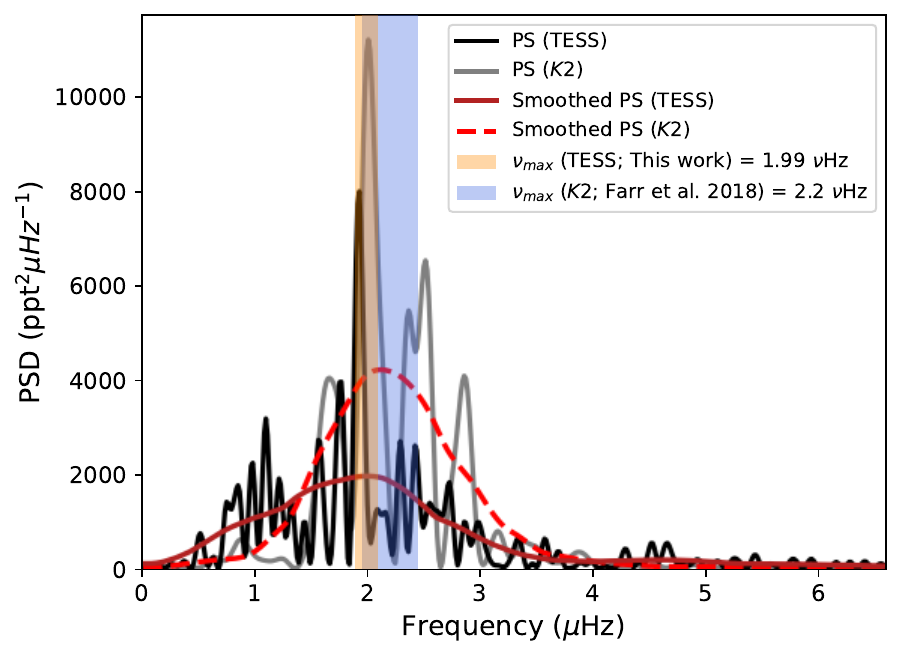}
 \caption{The power density spectrum of $\alpha$~Tau (Aldebaran) from both TESS (black) and $K$2 (gray). The smoothed spectra, which were smoothed with an Epanechnikov filter \citep{Epanechnikov1969}, with a width of $2 \ \mu \rm Hz$, are shown as a dark red solid line (TESS) and as a red dashed line ($K$2). The orange shaded region represents 1$\sigma$ region of the derived \numax{} from \textsc{pyMON}, while the blue shaded region represents 1$\sigma$ region of the $K$2 \numax{} from \citet{Farr2018}.}
 \label{fig:aldebaran_pymon}
\end{figure}

Aldebaran ($V$ = 0.87) is the brightest RG in our sample. The presence of solar-like oscillations in Aldebaran was confirmed by \citet{Farr2018} using RV variations from SONG-Tenerife, older archival observations, and halo photometry from $K$2. They derived a \numax{} value of 2.24$^{+0.09}_{-0.08}$\,$\mu$Hz and a mass of 1.16 $\pm$ 0.07\,$M_{\odot}$. Our measured \numax{} ($1.99 \pm 0.10 \, \mu \rm Hz$) and mass ($1.03 \pm 0.07 \, M_{\odot}$) are lower than their values, though our \numax{} is closer to their $K$2 value of 2.2 $\pm$ 0.25\,$\mu$Hz. In \autoref{fig:aldebaran_pymon} we show the power spectra from TESS and $K2$ for Aldebaran.

Earlier research suggested that Aldebaran hosted a planet with an orbital period of $\sim$629 days \citep{Hatzes1993, Hatzes1998b, Hatzes2015, Farr2018}. However, recent studies have raised doubts over this claim. A study by \citet{Hatzes2018} of $\gamma$~Dra, which has a long-term RV signal similar to that of Aldebaran, showed that its RV signal disappeared and reappeared with a phase shift. This behaviour would be inconsistent with an RV signal of planetary origin. Later, \cite{Reichert2019} reanalysed the RV data of Aldebaran with additional results from Lick Observatory, and they concluded that while a two-planet system fits the results better than one planet, it would be dynamically unstable. 

Aldebaran was observed by SONG, BRITE-Lem, and BRITE-Toronto to measure the amplitude ratio and phase shift between the asteroseismic signal in RV and in photometry \citep{Beck2020}, which coupled with the newly available TESS data could yield a more detailed asteroseismic analysis of Aldebaran. 

\subsection{$\beta$~Gem (Pollux)} \label{sec:ind_rg_stars_pollux}

Pollux ($V$ = 1.16) is a K0IIIb giant and is one of the brightest RGs in our sample. Oscillations were discovered by \citet{Hatzes2007} using 20~h of RV observations, and further confirmed by \citet{Hatzes2012} using longer RV time series and photometry from MOST. Like Aldebaran, Pollux was also thought to have an exoplanet companion \citep{Hatzes1993, Hatzes2006, Reffert2006}. However, Pollux was shown to be weakly magnetically active \citep{Auriere2009, Auriere2015}, with a periodicity similar to the claimed orbital period \citep{Auriere2021}, which might be the source of the RV signal. 

\begin{figure}
 \includegraphics[width=\columnwidth]{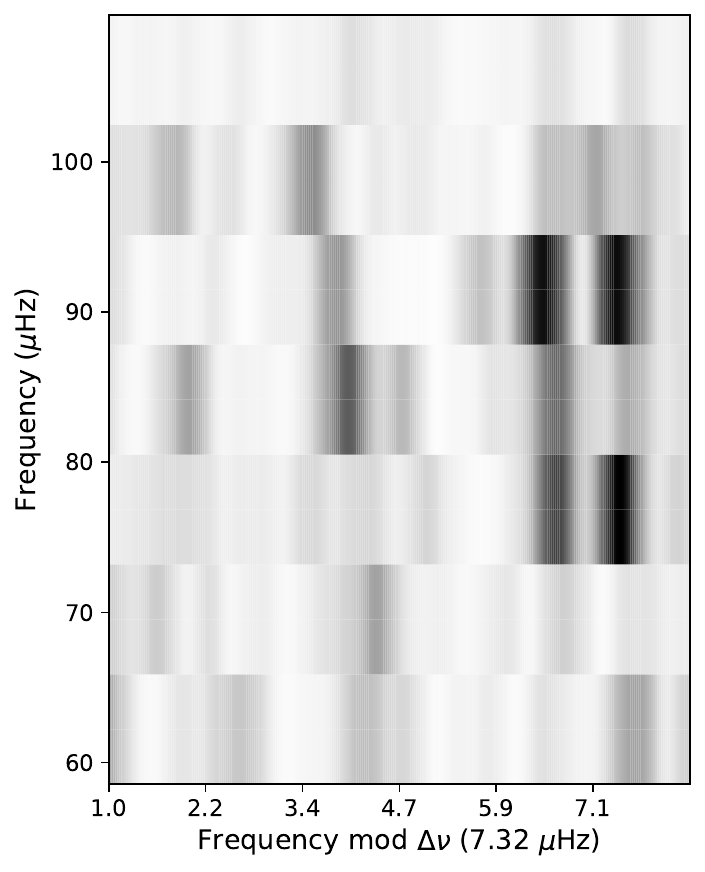}
 \caption{An \echelle{} diagram of $\beta$~Gem (Pollux) for \Dnu{} = $7.32$\,\muhz{}. The \echelle{} is shifted by an offset of 1\,\muhz{}, since the $\ell$ = 0 modes would otherwise be close to the right edge of the diagram.}
 \label{fig:pollux_echelle}
\end{figure}

Due to the high \numax{}, we were able to determine \Dnu{} using \echelle{} diagrams by aligning the $\ell = 0$ modes to be vertical, and with a value consistent with the empirical \Dnu{}–\numax{} relation by \citet{Yu2018a}. To validate our \Dnu{}, we also calculated the autocorrelation function (ACF) and confirmed a peak near the found value. Using both the \echelle{} method and the ACF, we found \Dnu{} for Pollux to be 7.32 $\pm$ 0.02\,\muhz{}. In \autoref{fig:pollux_echelle} we show the \echelle{} diagram of Pollux, in which we can see the $\ell = 0,2$ ridges located to the right, while the $\ell=1$ modes are located to the left.

Using both \numax{} and \Dnu{}, we calculated the masses and radii for this star using the standard scaling relations. For the correction factor on \Dnu{}, $f_{\Delta\nu}$ \citep{Sharma2016A}, we used Equation 16 by \citet{Li2023}, with [Fe/H] = 0.19 from \citet{AllendePrieto2004}. Using this equation yielded a seismic mass of $1.89 \pm 0.24\,M_{\odot}$ and a seismic radius of $8.47 \pm 0.3\newb{6}\,R_{\odot}$. The seismic radius is inconsistent with the interferometric radius of $8.97 \pm 0.03\,R_{\odot}$ \citep{Baines2025}. When we used no correction, we obtained a mass of $2.12 \pm 0.2\newb{6}\,M_{\odot}$ and radius of $8.97 \pm 0.3\newb{7}\,R_{\odot}$, which is more consistent with the interferometric radius. Using \autoref{eq:scaling_m_numax} we obtained a seismic mass of $2.12 \pm 0.09\,M_{\odot}$, which is consistent with the mass derived by not using $f_{\Delta\nu}$. If we used a different $f_{\Delta\nu}$ correction for red clump stars by \citet{Schimak2026}, we obtained a seismic mass of $2.09 \pm 0.2\newb{6}\,M_{\odot}$ and a seismic radius of $8.92 \pm 0.3\newb{7}\, R_{\odot}$, which are consistent with the interferometric radius and the mass from \autoref{eq:scaling_m_numax}. We list this radius and the corresponding mass for Pollux in \autoref{tab:gap}. While using this $f_{\Delta\nu}$ could suggest that Pollux is a red clump star, the position of Pollux in \autoref{fig:HR_astero} is also consistent with it ascending the red-giant branch.

Pollux has been extensively observed by SONG (see \autoref{tab:song_list} in \autorefA{appendix:song_list}) and is currently monitored to better constrain the origin of its long-term RV variability. These observations can be used with the available TESS data to analyse the star asteroseismically, with detailed modelling of its well-defined modes.

\subsection{$\alpha$~UMa (Dubhe)}

$\alpha$~UMa ($V$ = 1.81) is a binary system consisting of a K0III primary component \citep{Guenther2000} and a A5V secondary component \citep{Gray2018}. Solar-like oscillations were detected in the primary component by \citet{Buzasi2000b}, who used the $52$-mm star camera on WIRE \citep{Buzasi2000a}, making it one of the first detections of solar-like oscillations using space-based photometry. They claimed the detection of 10 modes, ranging from 1.82 to 43.56 $\mu$Hz, with a mode spacing of 2.94 $\mu$Hz. The oscillation we detected in the power spectrum, along with the derived \numax{} of $14.54 \pm 0.19$\,\muhz{} confirmed the earlier detection from WIRE. However, we do not detect modes below 9\,\muhz{} or above 18\,\muhz{}. A more in-depth study of the new \textsc{halophot} data and complementary SONG observations (see \autoref{tab:song_list} in \autorefA{appendix:song_list}) could validate or refute the modes reported by \citet{Buzasi2000b} and improve asteroseismic constraints for this star.

\subsection{$\beta$~UMi (Kochab)}

$\beta$~UMi ($V$ = 2.07) is a bright RG and, with 16 TESS sectors, has the second-most sectors in our sample. Oscillations in the star were detected by \citet{Tarrant2008} using SMEI. They used $\sim$1000 days of observation and were able to detect two modes at $2.44$ and $2.92$\,\muhz{}, which are consistent with our power spectrum. We also detected an additional mode around $3.37$\,\muhz{}, which explains why we obtained a \numax{} above 3\,\muhz{}. We note that $\beta$~UMi will be within 5$^\circ$ of \new{the proposed northern} PLATO field (see \autoref{tab:plato}).


\subsection{$\gamma$~Dra (Eltanin)}

This bright ($V$ = 2.24) low-\numax{} RG, with 18 TESS sectors, has the most sectors in our whole sample. We detected oscillations in this star with \numax{} of $2.92 \pm 0.12 \ \mu \rm Hz$. As mentioned in \autoref{sec:ind_rg_stars_aldebaran}, $\gamma$~Dra has a similar long-term RV variation as Aldebaran \citep{Hatzes1993, Hatzes2018}, which is unlikely to be of planetary origin \citep{Hatzes2018, Ramirez2020}. The star is inside PLATO's \new{proposed} northern field, within the field of view of its fast cameras (see \autoref{tab:plato}).

\subsection{$\alpha$~Cas (Schedar)}\label{sec:ind_rg_stars_alf_cas}

In $\alpha$~Cas ($V$ = 2.24) we detected power excess at 10\,\muhz{} in the star's power spectrum, which is the first reported detection of solar-like oscillations in the star. We report \numax{} \new{=} 9.68\,\muhz{} using \textsc{pyMON} and the derived mass using the scaling relation is 4.98 $\pm$ 0.40 $M_{\odot}$, which is unusually large for RGs \new{\citep{Crawford2024, Crawford2025}}. The only other study which derived the mass of $\alpha$~Cas was by \citet{Reffert2015}, who used stellar tracks from \citet{Girardi2000} to derive a mass of either 4.06 $\pm$ 0.35 or 3.98 $\pm$ 0.30\,$M_{\odot}$. These values are lower than our derived mass, but are still relatively high for RGs. A further study of $\alpha$~Cas is needed in order to confirm the mass of the star.

\subsection{$\beta$~Her (Kornephoros)}

The faintest star in our sample, $\beta$~Her ($V=2.78$), is a G7IIIa star, which we included because we noticed that the TPF did not encompass its bleed columns entirely. The star is a SB1 system \citep{Campbell1900, Plummer1908, Eaton2007,  Massarotti2008, Gray2016}, and its orbit was astrometrically resolved by \citet{Jancart2005} using Hipparcos data. If the projected semi-major axis and parallax are known, which is the case for $\beta$~Her, then it becomes possible to determine the mass sum and mass ratio of an astrometric SB1 binary system \citep[e.g.][]{Videla2022}. The star has been observed by CHARA/PAVO, enabling the determination of its angular diameter (Chowhan et al., in prep.) and is currently being monitored with SONG to constrain its orbit. Therefore, with asteroseismic independent mass and radius, $\beta$~Her promises to be an asteroseismic benchmark star in the near future. 

\section{Results for Pulsating A/F stars}\label{sec:res_af}

A-type and early F-type stars have \new{a very} thin \new{surface} convective \new{layer}, so the pulsations from these stars have a different origin than solar-like oscillators. In the classical instability strip we find the $\delta$~Sct variables, whose oscillations are driven by the $\kappa$-mechanism in the HeII ionisation zone \citep{Goupil2005, Handler2009} \new{and turbulent pressure \citep{Antoci2014, Antoci2019}}. 

\begin{figure*}
 \includegraphics[width=1.65\columnwidth]{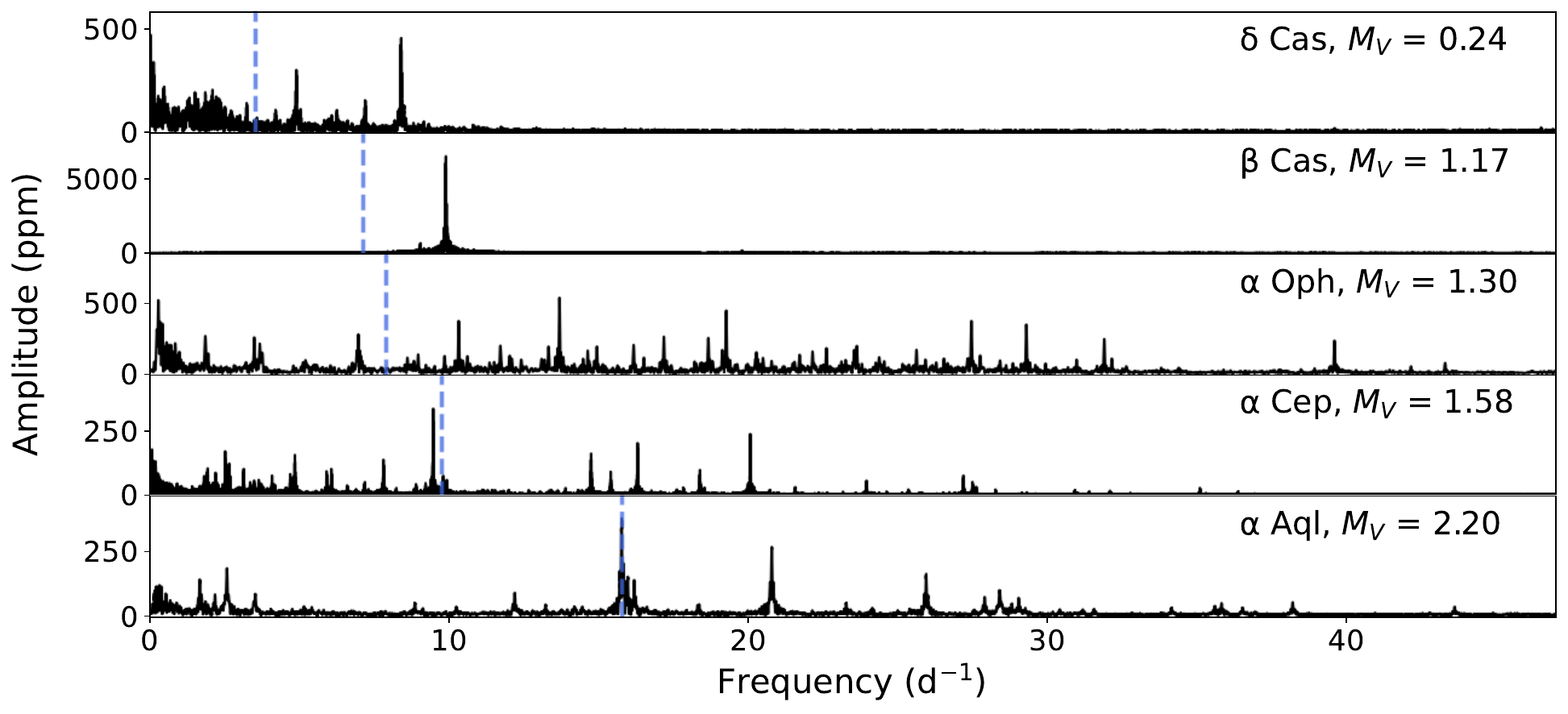}
 \caption{A gallery plot of the five $\delta$~Sct variables in our sampled sorted by their $M_V$. The excepted fundamental radial modes, calculated using luminosity-period relation from \citet{Barac2022}, are shown as blue dashed lines.}
 \label{fig:gallery_deltascu}
\end{figure*}

In our sample, we detected five $\delta$~Sct variables by examining their amplitude spectra, their spectral type, and previous literature. These are $\alpha$~Aql (Altair), $\alpha$~Oph (Rasalhague), $\beta$~Cas (Caph), $\alpha$~Cep (Alderamin), and $\delta$~Cas (Ruchbah). Their amplitude spectra are shown in \autoref{fig:gallery_deltascu}, along with their expected fundamental modes, which we calculated using the luminosity-period relation from \citet{Barac2022}. Among these five stars, we note that $\alpha$~Cep and $\delta$~Cas are newly discovered $\delta$~Sct variables. 

Overlapping with the red edge of the classical instability \new{strip} and extending to redder colours in the HR diagram are the $\gamma$~Dor variables, which are g-mode oscillators \new{whose oscillations are excited by flux blocking at the base of their thin outer convection zone \citep[e.g.][]{Guzik2000,Dupret2005}}. We detected only one clear $\gamma$~Dor variable, $\beta$~Ari ($V=2.64$), which was previously suspected to be a $\delta$~Sct variable \citep{BarbianodiBelgioso1983}. However, both $\alpha$~Oph and $\alpha$~Cep show signs of multiple low-frequency peaks as discussed in \autoref{sec:res_af_alf_oph} and \autoref{sec:res_af_alf_cep}. 

\new{Seven} stars in our sample are chemically peculiar Ap and Bp stars. \new{A small fraction of this class} are rapidly oscillating Ap stars (roAp), which exhibit rapid variability \new{\citep{Kurtz1990, Holdsworth2021}}. However, we did not detect any roAp signals in the 120-s cadence SPOC light curves among our Ap and Bp stars: $\alpha$~Eri (Achernar), \new{$\beta$~Tau \citep[Elnath;][]{Begari2026}}, $\epsilon$~UMa (Alioth), $\alpha$~And (Alpheratz), $\zeta$~UMa (Mizar), $\gamma$~Crv (Gienah), and $\theta$~Aur (Mahasim). For $\epsilon$~UMa, this confirms the earlier non-detection of oscillations by \citet{Retter2004}, who used observation from the 52-mm star camera on WIRE.


\subsection{$\alpha$~Aql (Altair)}
Altair ($V$ = 0.76) is the brightest known $\delta$~Sct variable, as discovered by \citet{Buzasi2005} using the 52-mm star camera on WIRE. Altair was then analysed by \citet{LeDizes2021} using the observations from MOST, and by \citet{Rieutord2024} using halo photometry on TESS Sector~54.

\subsection{$\alpha$~Oph (Rasalhague)}\label{sec:res_af_alf_oph}

The binary system, $\alpha$~Oph ($V = 2.08$), consists of a primary rapidly rotating A5IV star and a K5V-K7V star as its companion \citep{Hinkley2011}. A previous asteroseismic study of this system was done by \citet{Monnier2010} using 30 days of observation from MOST. They confirmed that the primary star is a $\delta$~Sct + $\gamma$~Dor hybrid. Our \textsc{halophot} amplitude spectrum (see \autoref{fig:rasalhague}) show clear detection of multiple peaks in the expected p- and g-mode frequency range, confirming the earlier detection. 

\begin{figure}
 \includegraphics[width=\columnwidth]{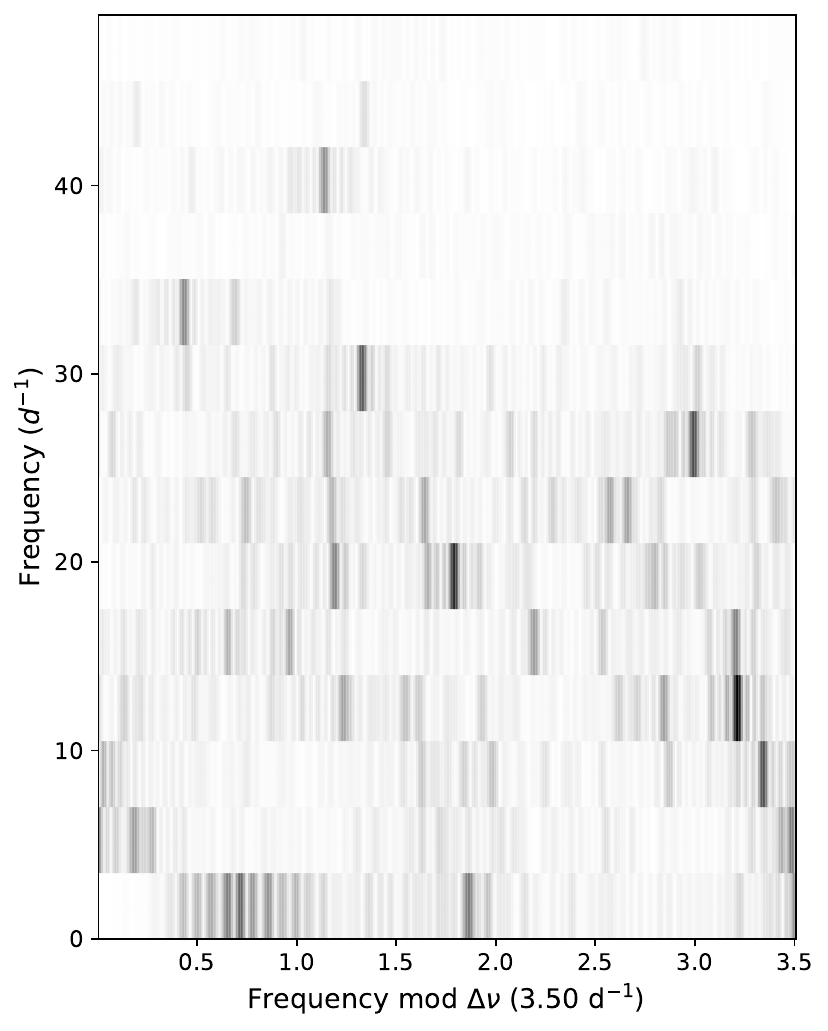}
 \caption{An \echelle{} diagram of $\alpha$~Oph (Rasalhague) with the most plausible value of \Dnu{}.}
 \label{fig:rasalhague_echelle}
\end{figure}

We attempted to find \Dnu{} for this star using the \echelle{} diagram by aligning the $\ell = 1$ modes to be vertical \citep{Bedding2020}, since we know the dynamical mass and interferometric radius of the oscillating component. We found three possible values of \Dnu{}: 3.35, 3.37, and 3.50\,d$^{-1}$. We chose 3.50\,d$^{-1}$ as the most plausible value and present its corresponding \echelle{} diagram in \autoref{fig:rasalhague_echelle}. We compared this value to the theoretical \Dnu{} calculated from the star's density, using the dynamical mass ($2.20\,M_{\odot}$) from \citet{Gardner2021} and the interferometric radius ($2.623\,R_{\odot}$)\footnote{Calculated as the average between the equatorial ($2.858\,R_{\odot}$) and polar radius ($2.388\,R_{\odot}$)} from \citet{Monnier2010}. We obtained a theoretical \Dnu{} of 3.44 d$^{-1}$, which matches the three possible \Dnu{} values we found.

\subsection{$\beta$~Cas (Caph)}
$\beta$~Cas ($V$ = 2.28) was discovered to be variable by \citet{Millis1966}, making it the third brightest $\delta$~Sct variable in our sample. Unlike the other $\delta$~Sct variables in our sample, $\beta$~Cas has relatively few peaks in its amplitude spectrum. \citet{Zwintz2020} analysed $\beta$~Cas using observations from BRITE, SMEI and TESS, along with spectropolarimetry observations, and found evidence that the star has a dynamo magnetic field. 

\subsection{$\alpha$~Cep (Alderamin)}\label{sec:res_af_alf_cep}

A bright ($V$ = 2.45) A7IV-V star \citep{Belle2006}, $\alpha$~Cep was suspected to be a $\delta$~Sct variable by \citet{Hauck1971}. It was therefore listed as a possible variable in the General Catalogue of Variable Stars \citep[GCVS\footnote{\url{http://www.sai.msu.su/gcvs/gcvs/}};][]{Samus2017}, and as a $\delta$~Sct variable in The International Variable Star Index \citep[VSX\footnote{\url{https://www.aavso.org/vsx/index.php}};][]{Watson2016}, but apparently without observations of variability. The star was observed by BRITE \citep{Zwintz2024}, but they did not detect any variability. However, all 11 TESS sectors of this star show small-amplitude peaks characteristic of a $\delta$~Sct star, making it the brightest $\delta$~Sct variable discovered by TESS. 

We also see multiple peaks below the expected frequency of the  fundamental radial mode ($\sim$9.7 d$^{-1}$; see \autoref{fig:gallery_deltascu}). This value suggests classification of the star as a $\delta$~Sct + $\gamma$~Dor hybrid. Furthermore, the star will be inside PLATO's \new{proposed} northern field (see \autoref{tab:plato}), and therefore has the potential to be one of the best-studied bright $\delta$~Sct--$\gamma$~Dor hybrids in the future.

\subsection{$\delta$~Cas (Ruchbah)}

A newly discovered $\delta$~Sct variable, $\delta$~Cas ($V=2.66$), is the second faintest star in our total sample. It was initially classified as an EB, with a period of 759~d, in the GCVS and was subsequently classified as such in Catalogue of Eclipsing Variables \citep{Malkov2006, Avvakumova2013} and VSX \citep{Watson2016}. However, we did not see any eclipses in either the \textsc{halophot} or SPOC light curves. It was observed by BRITE \citep{Zwintz2024}, but they did not detect any variability. This does not rule out the possibility that the star is eclipsing due to the claimed orbital period; however, we note that the current GCVS lists the star as a non-variable \citep{Samus2017}.

\section{Results for Pulsating O/B stars}\label{sec:res_ob}

Some O and B stars pulsate as $\beta$~Cep variables, where the $\kappa$-mechanism drives the p- and g-mode oscillations in the iron opacity bump at $\sim$200,000 K \citep{Dziembowski1993a, Thoul2009}. Of the stars in our sample, nine are $\beta$~Cep variables, including the ellipsoidal (ELL) variable $\alpha$~Vir \new{\citep[Spica;][]{Tkachenko2016}} and the EB $\lambda$~Sco (Shaula).

Cooler B-type stars can also be SPB stars, which are g-mode pulsators \citep{Dziembowski1993b}. We detected three SPB stars and one candidate, $\epsilon$~Car, which might be an LPV + SPB asteroseismic binary (see \autoref{sec:ind_OB_stars_eps_car}). 
\new{Additionally, Be-type stars have previously been found to exhibit photometric variability \citep[e.g.][]{Gutierrz-Soto2007,Labadie-Bartz2017,Labadie-Bartz2022, Rivinius2026}
. These are non-supergiant B-type stars that have or at some point had Balmer lines in emission \citep{Collins1987,Rivinius2013}, which has been attributed to the presence of a circumstellar disk \citep[e.g.][]{Struve1931,Carciofi2009}.} 
We classified eight stars in our sample as variable Be stars, including the brightest star, $\alpha$~Eri (Achernar).

\begin{figure*}
 \includegraphics[width=1.65\columnwidth]{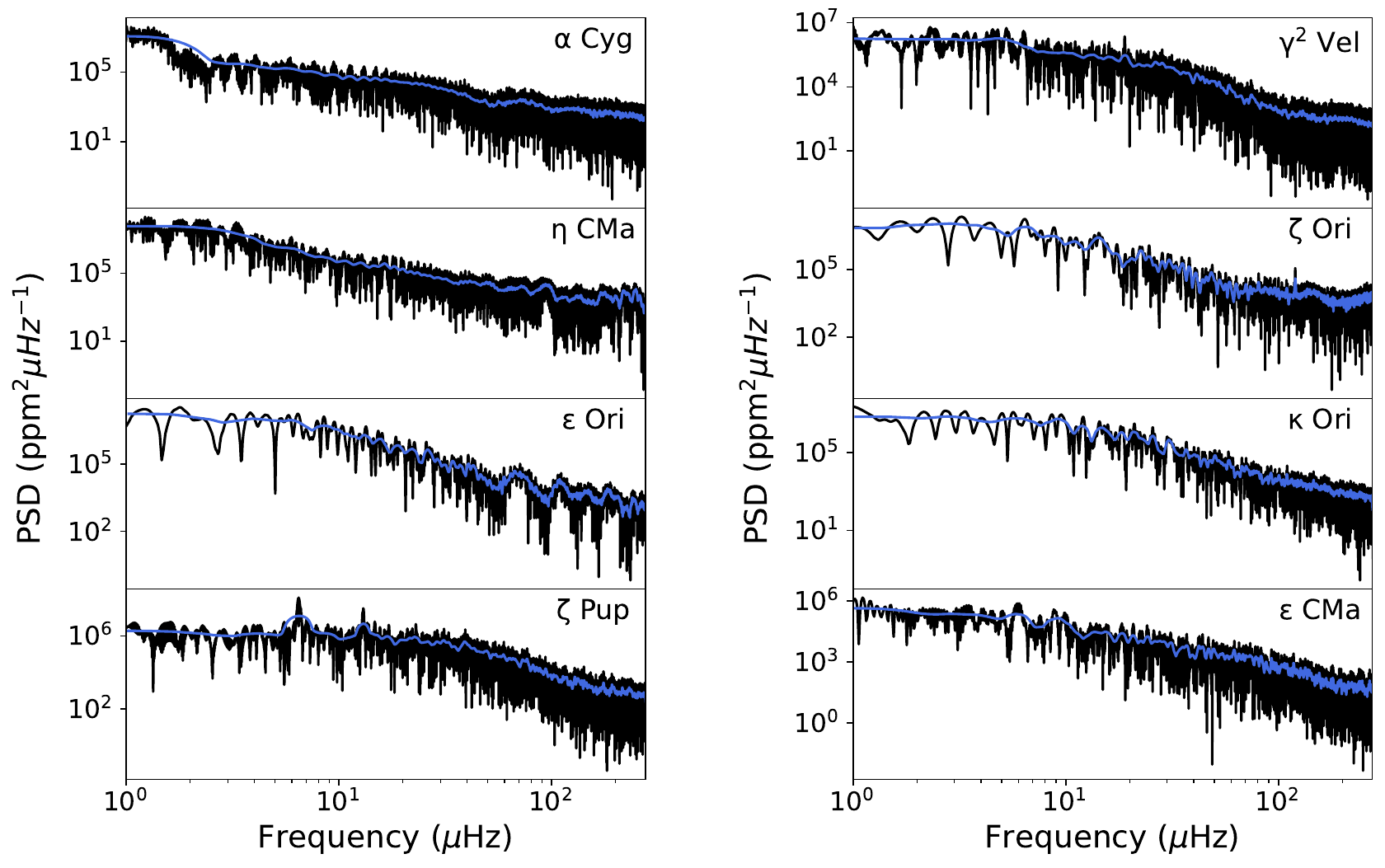}
 \caption{A gallery plot of the 8 SLFs in our sample. The weighted power spectral densities are shown in black. The smoothed spectra with an Epanechnikov filter \citep{Epanechnikov1969}, with a width of $2$\,\muhz{}, are shown in blue.}
 \label{fig:slfs}
\end{figure*}

Another form of variability observed in O and B stars is stochastic low-frequency (SLF) variability. These SLF variables are characterised in the frequency domain by a stochastic signal at low frequencies that decreases in power towards higher frequencies. 
\new{The SLF variability was initially characterised for three O-type stars by \citet{Blomme2011} and was later found to be ubiquitous in massive stars, including main-sequence O- and B-type stars \citep[e.g][]{Bowman2019b,Bowman2019,Bowman2020,Pedersen2025}, blue supergiants \citep[e.g.][]{Bowman2019b,Bowman2020,Kourniotis2025}, yellow and red supergiants \citep{Dorn-Wallenstein2019,Zhang2024}, and Wolf-Rayet stars \citep[e.g.][]{Lenoir-Craig2022}.}
In our sample, we classified eight stars as SLFs and one star, $\theta$~Sco, as a potential SLF candidate (see \autoref{sec:ind_rg_stars_tet_sco}). 
We show their weighted power spectral densities in \autoref{fig:slfs}. 

\subsection{$\gamma^2$~Vel}\label{sec:ind_OB_stars_gam_vel}

\begin{figure}
 \includegraphics[width=\columnwidth]{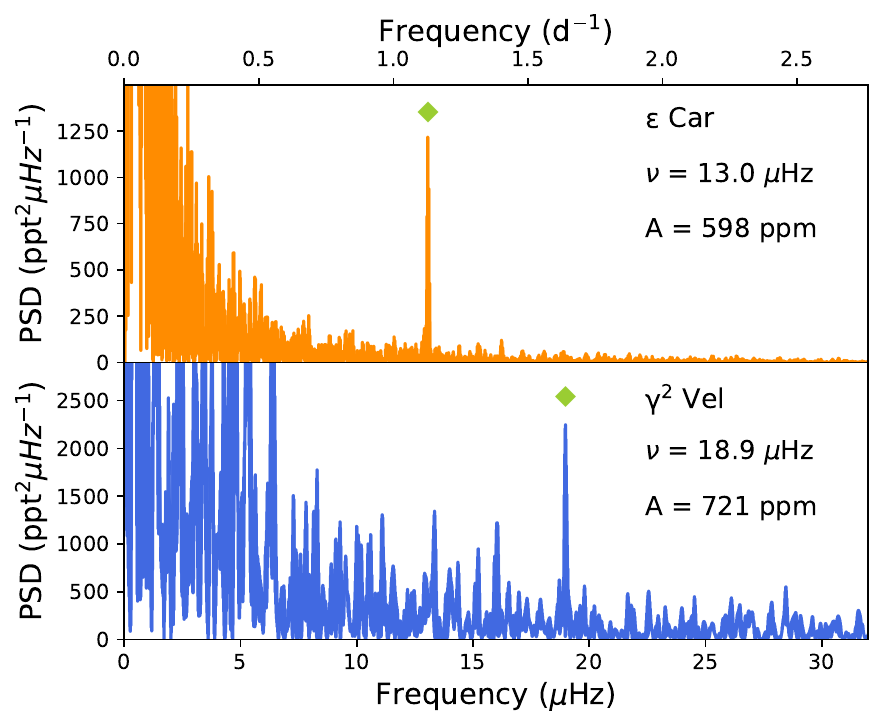}
 \caption{Power density spectra for $\epsilon$~Car (top) and $\gamma^2$~Vel (bottom). Their pre-whiten frequencies are indicated by the green diamonds, with their corresponding frequencies ($\nu$) and amplitudes (A) shown on the right side of the plot.}
 \label{fig:clean}
\end{figure}

$\gamma^2$~Vel ($V$ = 1.75) is a SB2 system consisting of a WC8 Wolf-Rayet star and an O9I supergiant \citep{Hoffleit1995}, with an orbital period of 78.53~d \citep{1997Schmutz}. Variability was detected by \citet{Richardson2017} using BRITE observations and 488 high-resolution optical spectra, and was attributed to colliding winds. We detected variability characteristic of SLF, potentially making it the brightest known Wolf–Rayet star with stochastic variability. The variability detected by \citet{Richardson2017} had a longer timescale than the SLF signal we observed. However, the BRITE amplitude spectrum (Fig. 5 in \citealt{Richardson2017}) appears to show characteristics similar to 
\new{the TESS SLF signal}. We also detected a peak at 18.94\,\muhz{}, measured using sine-wave fitting \citep{Roberts1987}, which we show in \autoref{fig:clean}.

We note that $\gamma^1$~Vel ($V$ = 4.21) is only 41" from $\gamma^2$~Vel \citep{Hernandez1980}, and could thus contaminate the light curves. A future study incorporating the TESS, BRITE, and PLATO (see \autoref{tab:plato}) observations would help characterise the stochastic nature of this system even further. However, the question of saturation for $\gamma^1$~Vel and $\gamma^2$~Vel in PLATO could complicate the photometry. 

\subsection{$\theta$~Sco (Sargas)}\label{sec:ind_rg_stars_tet_sco}

$\theta$~Sco ($V=1.86$) is binary system consisting of either a F1II \citep{Hoffleit1995} or a F1III \citep{Gray1989} primary component, along with a $\sim$A3 giant secondary component \citep{Lewis2022}. The primary component is a rapid rotator, and could therefore be a post-merger object \citep{Lewis2022}. \new{BRITE flagged the star as a variable \citep{Zwintz2024} and we} detected SLF-like variability in both TESS sectors. 

\subsection{$\epsilon$~Car (Avior)}\label{sec:ind_OB_stars_eps_car}

A long-period binary system consisting of a K3III primary component and a B2:V secondary component \citep{Hoffleit1995}, which are 0.41" apart \citep{Parsons1998}, $\epsilon$~Car ($V=1.86$) was flagged by BRITE as a variable \citep{Zwintz2024} \new{and the secondary component was previously analysed by \citet{Bowman2022} using TESS, but they ruled out $\beta$~Cep pulsations in the star}. From the \textsc{halophot} light curves, all 12 TESS sectors revealed a frequency signal at 13~\muhz{} and a low frequency background signal, which we show in \autoref{fig:clean}. The frequencies, along with their spectral types, could point to $\epsilon$~Car being an LPV + SPB asteroseismic binary. Alternatively, the companion was classified as a B2Vp star by \citet{Parsons1998}, suggesting that it could be an $\alpha^2$~CVn variable. Finally, we note that this system is within PLATO's southern field (see \autoref{tab:plato}), \new{potentially} providing a long time series for detailed asteroseismology.

\subsection{$\sigma$~Sgr (Nunki)}

$\sigma$ Sgr ($V$ = 2.05) is an interferometric binary \citep{Hanbury_Brown1974, Bedding1994}. \new{The star was analysed by \citet{Telting2006} using ground-based spectrographs and later by \citet{Bowman2022} using TESS for $\beta$~Cep pulsations, but neither studies detected any sign of variability.} The light curve for this star showed some variability and multiple peaks in the amplitude spectrum below 3 d$^{-1}$, although with only one sector, the peaks are not resolved enough. We classified the star as a SPB variable. 

\section{Results for Eclipsing binaries}\label{sec:res_eb}

If the orientation of two binary components aligns such that one or both produce periodic dips in the light curve, then the system is classified as an EB. The depth and shape of the eclipses depend on the relative size of the components, their temperatures, and the orientation of their orbit. EBs observations can be complemented by RV observations to determine model-independent masses and radii for both components. The derived masses and radii can be used to test stellar models \citep[e.g.][]{Higl2017} and, in a few cases, combined with asteroseismic results \new{\citep[e.g.][]{Gaulme2016, Li-Tanda2018, Benbakoura2021, Brogaard2022, Gaulme2022, Thomsen2025, Southworth2025}}. 

Of the stars in our sample, 8 are EBs. Except for $\gamma$~And, all were already known to be eclipsing, with $\lambda$~Sco being the most recent discovered EB \citep{Buzasi2004, Bruntt2006}. In \autoref{tab:list_ebs} we list the EBs in our sample, along with the eclipsing components, the periodicity in days, and individual notes for some of them. 
\begin{table*}
	\centering
	\caption{List of EBs in our sample.}
	\label{tab:list_ebs}
	\begin{tabular}{cccccc}
		\hline
		Star & $V$ & Eclipsing components & Orbital period (d) & \new{Period reference} & Notes \\
		\hline
		$\alpha$~Gem & 1.58 & Ca + Cb & 0.814 & \new{\citet{Torres2002}} & \\ 
        $\lambda$~Sco & 1.62 & Aa1 + Aa2 & 5.925 & \new{\citet{Tokovinin2018}} & $\beta$~Cephei variable. Observed by WIRE$^1$ \\
        $\beta$~Aur & 1.90 & Aa + Ab & 3.960 & \new{\citet{Southworth2007}} & Observed by WIRE$^2$ and BRITE$^3$ \\ 
        $\delta$~Vel & 1.93 & Aa + Ab & 45.150 & \new{\citet{Merand2011}} & Observed by SMEI$^4$ and BRITE$^5$ \\
        $\beta$~Per & 2.09 & A + B & 2.867 & \new{\citet{Baron2012}} & Observed by BRITE$^5$ \\
        $\gamma$~And & 2.10 & Ba + Bb & 2.670 & \new{\citet{Maestre1960}} & See  \autoref{sec:ind_EB_stars_almach} \\ 
        $\alpha$~CrB & 2.22 & A + B & 17.360 & \new{\citet{Tomkin1986}} &  \\
        $\delta$~Ori & 2.25 & Aa1 + Aa2 & 5.732 & \new{\citet{Mayer2010}} & Tidally pulsating. Observed by MOST$^{6}$ and BRITE$^5$ \\ 
		\hline
        \multicolumn{6}{l}{1: \citet{Bruntt2006}, 2: \citet{Southworth2007}, 3: \citet{Strassmeier2020}, 4: \citet{Pribulla2011}, 5: \citet{Zwintz2024}, 6: \citet{Pablo2015}} \\
	\end{tabular}
\end{table*}




\subsection{$\beta$~Per (Algol)} 
Algol ($V=2.09$) is one of the first known variable stars \citep{Montanari1671}. \citet{Goodricke1783} proposed that the variability could be caused by an object eclipsing the star, and the binary nature of the system was confirmed by \citet{Vogel1890}, who resolved the star as a spectroscopic binary. 

There have been many photometric observations of Algol \citep[e.g.][]{Stebbins1910, Chen1966, Smyth1975, Eaton1975, Longmore1975, Zeilik1980, Al-Naimiy1985, Kim1989}, and so the eclipse depth of this star is well known across a broad wavelength range. We therefore use Algol to validate the amplitude of the \textsc{halophot} light curves in \autoref{sec:discussion_amplitude}.

\subsection{$\gamma$~And (Almach)}\label{sec:ind_EB_stars_almach}
$\gamma$~And ($V$ = 2.10) is a multiple system consisting of five stars. The brightest star, $\gamma^1$ And, is a K2+IIb giant \citep{Keenan1989}, which is 9.9" separated from its companion $\gamma^2$ And, which itself is an interferometric system consisting of $\gamma$~And B and C. The B component was discovered to be an SB2 by \citet{Maestre1960}, who derived an orbital period of 2.670 days and an eccentricity of 0.292. The \textsc{halophot} light curves reveal an eclipsing binary with an orbital period of $\sim$2.6 days, \new{with secondary eclipses that are offset from the midpoints of the primary eclipses, showing that the system is eccentric.} \new{These properties of the eclipsing binary are} consistent with the orbital parameters reported by \citet{Maestre1960}. We conclude that the B component is a newly discovered EB. Furthermore, the orbits of the B and C were resolved using speckle interferometry \citep{Docobo2007}, making it one of the few systems where individual masses and radii, along with the distance, can be dynamically determined.



\section{Comparison of TESS halo photometry with other telescopes}\label{sec:discussion}

\subsection{Comparison with the $K$2 Bright Star Survey}\label{sec:discussion_K2}

The stars $\alpha$~Tau (Aldebaran; see \autoref{sec:ind_rg_stars_aldebaran}), $\alpha$~Vir (Spica), $\delta$~Sco (Dschubba), and $\zeta$ Sgr (Ascella) are in the $K$2 Bright Star Survey \citep{Pope2019}, which also used \textsc{halophot}. Comparing the \textsc{halophot} results between this work and $K$2 Bright Star Survey enables us to test the consistency between the TESS and $K$2 results and compare the four stars across different instruments.

\begin{figure}
 \includegraphics[width=\columnwidth]{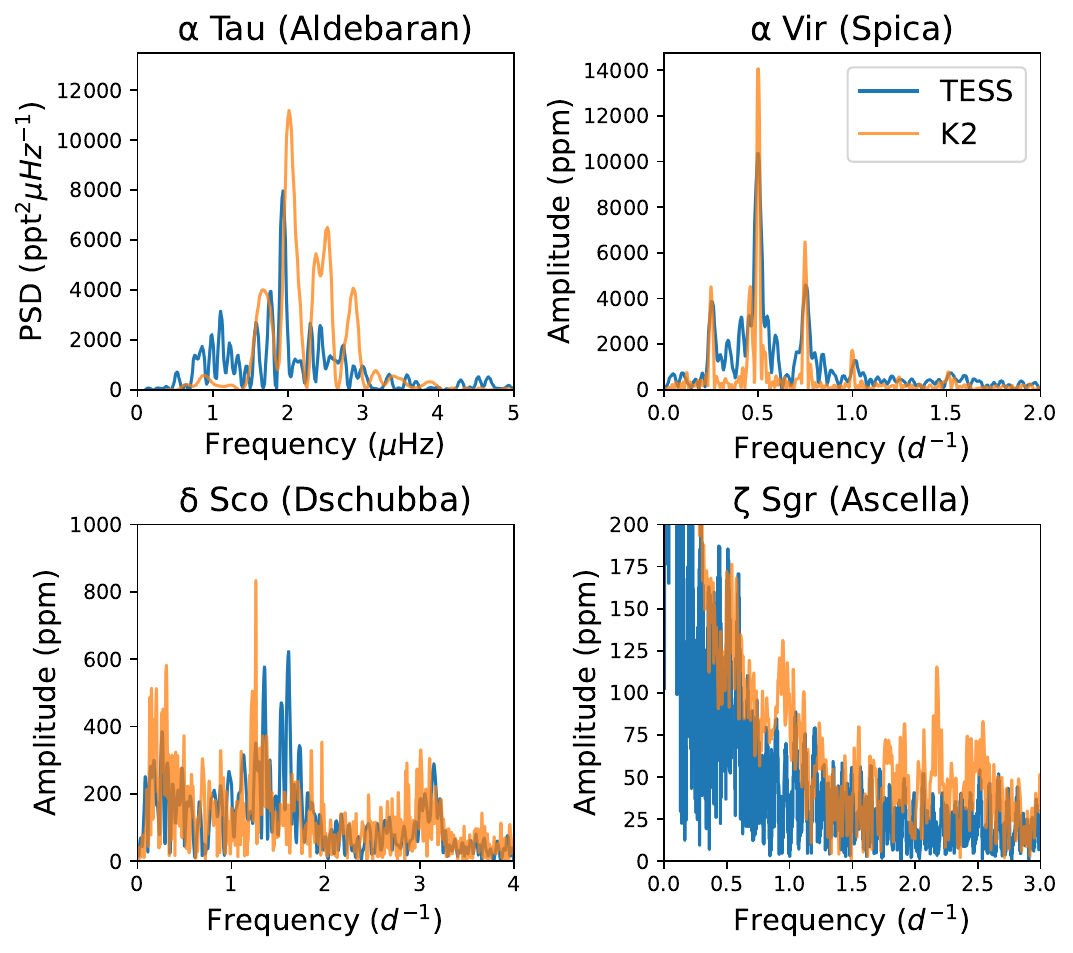}
 \caption{A comparison plot for the power spectrum of the four stars in the $K$2 Bright Star Survey. The blue power spectrum is from TESS, while the orange is from $K$2. }
 \label{fig:k2_comparision}
\end{figure}

We downloaded the $K$2 \textsc{halophot} light curves from MAST, and de-trended them using the same method as described in \autoref{sec:extract_lcurves}. We computed the weighted power spectra, which are shown in \autoref{fig:k2_comparision} along with the TESS \textsc{halophot} weighted power spectra.

For Aldebaran and $\delta$~Sco, the peaks are not precisely aligned\new{. For Aldebaran, we attribute the non-aligned signal to the star being a stochastic oscillator. For $\delta$~Sco, the non-aligned signal is due to the star being a Be star, which can either have resolved peaks but with variable amplitudes, or the beating of unresolved coherent pulsations can cause time dependency on the light curve. Since both $K$2 and TESS observed both stars at different times, it is not surprising that the peaks of Aldebaran and $\delta$~Sco are not well aligned.}

For Spica, the peaks are well aligned in the amplitude spectra, since the star is a coherent variable \citep[ELL + $\beta$~Cep;][]{Tkachenko2016}. The amplitude from $K$2 is slightly larger than that from TESS, which may be due to differences between the TESS and \kepler{} passbands \citep{Lund2019}. The signal is also better resolved in $K$2 because the star was observed in only one TESS sector ($\sim$27 d), compared to the one campaign in $K$2 ($\sim$80 d).

The star $\zeta$ Sgr does not show variability in TESS (in either \textsc{halophot} or SPOC), but the $K$2 amplitude spectrum shows a peak around $\sim$2.1\,d$^{-1}$, and the star was listed as a $\gamma$~Dor variable by \citet{Pope2019}. 
The $K$2 light curve for this star contains strong trends, so the claimed $\gamma$~Dor signal seen in $\zeta$ Sgr could be due to instrumental systematics.

\begin{figure}
 \includegraphics[width=\columnwidth]{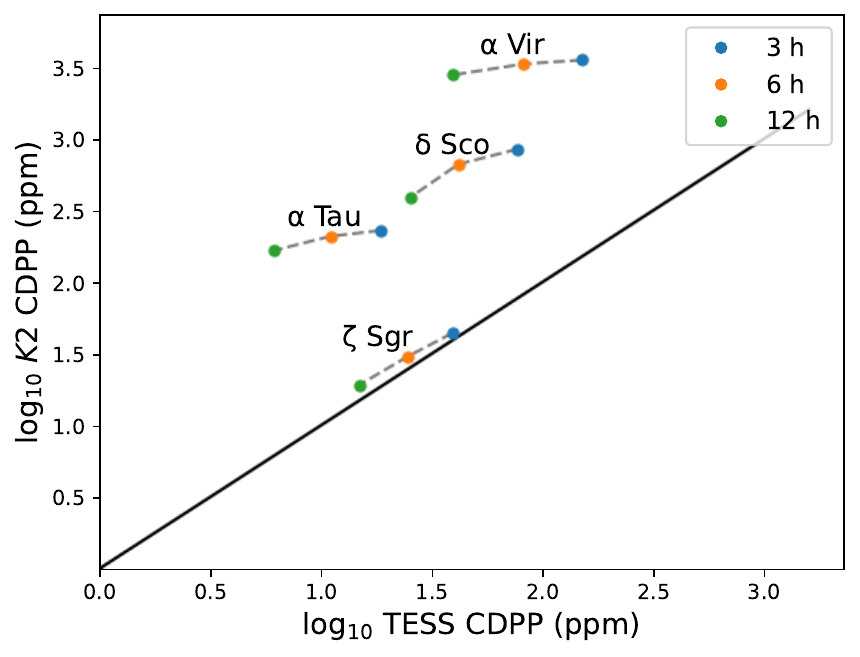}
 \caption{Correlation diagram of the 3, 6, and 12 h \textsc{lightkurve}-computed CDPP for both the TESS and $K$2 \textsc{halophot} light curves. Grey dashed lines connect the CDPPs associated with a given star.}
 \label{fig:cdpp_comparision}
\end{figure}

We also compared the quality of the light curve using 3, 6, and 12~h Combined Differential Photometric Precision \citep[CDPP;][]{Christiansen2012} as our proxy for noise. We computed the CDPP using \textsc{lightkurve} for both the TESS and $K$2 \textsc{halophot} light curves. For the TESS light curves, we took the mean CDPP for each sector as our comparison value. We show the results in \autoref{fig:cdpp_comparision}. From the figure, the \textsc{halophot} light curves from $K$2 have larger CDPP values, meaning that the noise is generally lower for the TESS data than the $K$2 data for this small sample. 

\subsection{Validity of halo photometry amplitudes with eclipsing binaries}\label{sec:discussion_amplitude}

Well-calibrated amplitudes are important for EB studies, and also for simultaneous RV and photometry studies. Therefore, we did a rough analysis in order to test the \textsc{halophot} amplitude. We chose the EBs, $\beta$~Aur (Menkalinan), $\beta$~Per (Algol), and $\alpha$~CrB (Alphecca), since they have the best light curves, no significant intrinsic stellar variability, and no long-term trend in their light curves. We calculated the eclipse depth for each sector by subtracting the median flux from the maximum flux, and took the mean of each sector's eclipse depth to obtain the primary eclipse depth for the given star. We compared these values with the eclipse depths reported in the literature, which we estimated from published light curves. We chose light curves obtained with a photometric filter that overlaps the TESS detector's bandpass, which spans 600--1000~nm. 

For $\beta$~Aur, the primary eclipse depth from \textsc{halophot} was 86.6~mmag, which matches with $\sim$80~mmag from BRITE-Toronto, which has a wavelength range of about 550--700~nm \citep{Strassmeier2020}. Algol has an eclipse depth with \textsc{halophot} of 1059~mmag, which lies between the two eclipse depths of $\sim$925~mmag and $\sim$1150~mmag from \citet{Chen1966}, who used filters centred at 1000~nm and 600~nm, respectively. Lastly, $\alpha$~CrB has an eclipse depth with \textsc{halophot} of 114~mmag, which is consistent with the depth of $\sim$100~mmag reported by \citet{Kron1953} using a filter centred at 723 nm. We also calculated the eclipse depth from the SPOC and QLP light curves, deriving values of 53~mmag ($\beta$~Aur), 562~mmag (Algol), and 98~mmag ($\alpha$~CrB)\footnote{We ignored TESS Sectors 51 and 78 due to the QLP light curves containing a lot of noise}, which were all much lower than literature values, which we attribute to some (but not all) the saturated pixels being inside the aperture mask (see \autoref{sec:discussion_quality}). The \textsc{halophot} eclipse depths for all three EBs match the literature, giving confidence in the \textsc{halophot} amplitudes. 

\section{Future work with the TESS halo photometry sample}\label{sec:future_work}


\subsection{PLATO}\label{sec:future_work_plato}

We identified 18 stars in our sample that will be observed by the upcoming PLAnetary Transits and Oscillations of stars (PLATO) mission \citep{Rauer2025}, scheduled to launch in \newb{early 2027}. PLATO has 24 cameras with 25-s cadence, plus two additional fast cameras with 2.5-s cadence that will provide blue (505--700 nm) and red (665--1000 nm) photometry.

\begin{figure}
 \includegraphics[width=\columnwidth]{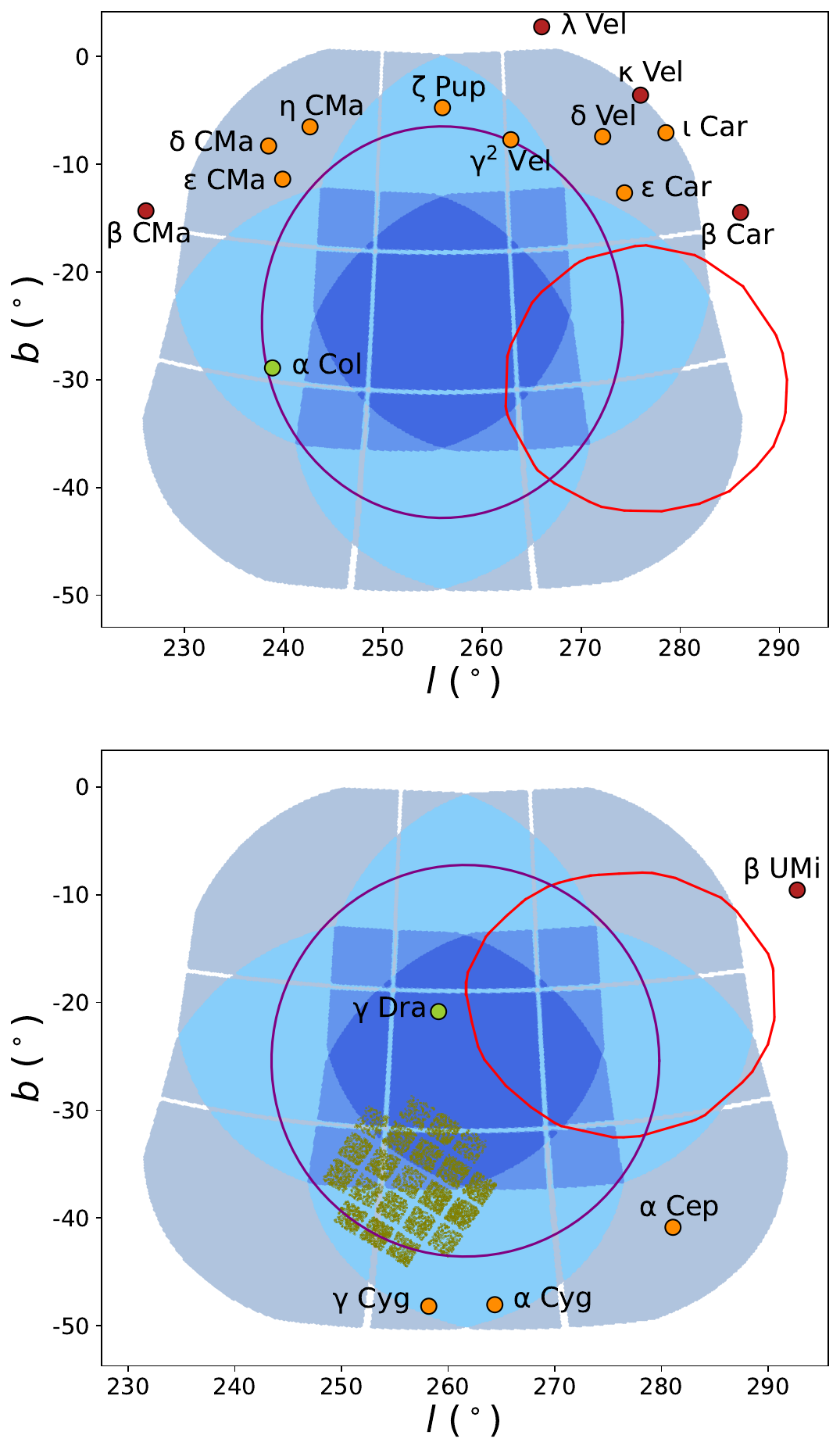}
 \caption{The two PLATO LOPs with the stars in our sample. \new{Top} panel shows the PLATO LOPS2 with the southern stars. \new{Bottom} panel is the \newb{proposed} PLATO LOPN1 with the northern stars. In both panels the LOPs are shown blue, with the different shade indicating either 24, 18, 12, or 6 overlapping cameras. The red circles are the TESS CVZs, while the \kepler{} field is shown in olive-green. The field of views of the fast cameras are shown in purple. Orange are stars inside the PLATO LOP, while red are stars within 5$^\circ$ of the LOP. The green are stars inside the field of view of the fast cameras.}
 \label{fig:plato}
\end{figure}

\newb{The current plan for PLATO is to observe a LOP (Long-duration Observation Phase) field in the south and potentially one LOP field in the north.} These LOPs will each cover a 49$^\circ$ $\times$ 49$^\circ$ region of the sky, overlapping with the two TESS CVZs and the \kepler{} field \citep{Rauer2025}. PLATO's current strategy is to observe the southern LOP for two years. \newb{Afterwards, it will either continue observing the southern LOP for another two years or shift its field of view to observe the northern LOP for two years}. At the time of writing, only the southern LOP (LOPS2), centred around $\alpha = 95.31043$, $\delta = -47.88693$, is defined \citep{Nascimbeni2025}, while for the northern LOP we adopt the candidate field, LOPN1, centred around $\alpha = 277.18023$, $\delta = 52.85952$ \citep{Nascimbeni2022}. In \autoref{fig:plato} we show the LOPs in Galactic coordinates, along with the stars in our sample. 

The stars in our sample that PLATO \newb{could} observe are listed in \autoref{tab:plato}. Of these, nine will be inside LOPS2, while four stars \newb{are within} LOPN1. Furthermore, five stars \newb{could} be within 5$^\circ$ of the two LOPs and so may be observed if the fields are shifted. All of our stars fall outside the dynamic range of the fast cameras (4 < $V$ < $8$) and the optimised range of the normal cameras ($V \geq$ 8), so they will saturate the detector. Furthermore, the stars in our sample do not meet the requirements to be included in the P1, P2, P4, or P5 samples \citep{Montalto2021, Rauer2025}, but could be included as complementary science targets. Two stars, $\alpha$~Col and $\gamma$~Dra, \newb{might} meet the requirements of the Colour Sample, as they \newb{could be} within the nominal field of view of the fast cameras, allowing imagettes (equivalent to TPFs) to be provided. However, these stars may not be observed because the effective field of view is limited to $\sim$610 deg$^2$ \citep{Jannsen2024, Rauer2025} due to the use of frame-transfer CCDs in the fast cameras. For the rest of the sample, it is unclear whether photometry can be performed on board the spacecraft, although halo photometry is being considered for highly saturated stars \citep{Rauer2025}. Our results from TESS can be used to guide the selection of stars for PLATO observations.
\begin{table}
	\centering
	\caption{Stars in the PLATO fields}
	\label{tab:plato}
	\begin{tabular}{lcccccc}
		\hline
		Star & LOP & RA & DEC & Sep & $V$ & Variability \\
         & & (deg) & (deg) & (deg) & (mag) & \\
        \hline
        & \multicolumn{5}{c}{Inside the fast cameras' field of view} & \\
        \hline
        $\gamma$~Dra & N1 & 269.15 & 51.49 & 5.11 & 2.24 & RG \\
        $\alpha$~Col & S2 & 84.91 & -34.07 & 15.85 & 2.65 & Be \\
        \hline
        & \multicolumn{5}{c}{Inside PLATO LOPs} & \\
        \hline
		$\alpha$~Cyg & N1 & 310.36 & 45.28 & 22.78 & 1.25 & SLF \\
        $\epsilon$~CMa & S2 & 104.66 & -28.97 & 20.25 & 1.50 & SLF \\
        $\gamma$$^2$~Vel & S2 & 122.38 & -47.34 & 18.17 & 1.75 & SLF \\
        $\delta$~CMa & S2 & 107.10 & -26.39 & 23.39 & 1.83 & \\
        $\epsilon$~Car & S2 & 125.63 & -59.51 & 21.10 & 1.86 & LPV + SPB/ \\
        & & & & & & $\alpha^2$~CVn \\
        $\delta$~Vel & S2 & 131.18 & -54.71 & 23.15 & 1.93 & EB \\
        $\iota$~Car & S2 & 139.27 & -59.28 & 27.83 & 2.21 & \\
        $\zeta$~Pup & S2 & 120.90 & -40.00 & 19.92 & 2.21 & SLF \\
        $\gamma$~Cyg & N1 & 305.56 & 40.26 & 22.99 & 2.23 &  \\
        $\eta$~CMa & S2 & 111.02 & -29.30 & 22.18 & 2.45 & SLF \\
        $\alpha$~Cep & N1 & 319.64 & 62.59 & 24.12 & 2.45 & $\delta$~Sct + $\gamma$~Dor \\
		\hline
        & \multicolumn{5}{c}{Near PLATO LOPs} & \\
        \hline
        $\beta$~Car & S2 & 138.30 & -69.72 & 154.88 & 1.67 & Var \\
        $\beta$~CMa & S2 & 95.67 & -17.96 & 29.93 & 1.98 & $\beta$~Cep \\
		$\beta$~UMi & N1 & 222.68 & 74.16 & 145.56 & 2.07 & RG \\
        $\lambda$~Vel & S2 & 137.00 & -43.43 & 29.11 & 2.23 & LPV \\
        $\kappa$~Vel & S2 & 140.53 & -55.01 & 28.52 & 2.23 & SPB \\
		\hline
	\end{tabular}
\end{table}
\subsection{SONG}\label{sec:future_work_song}

The Stellar Observations Network Group (SONG) is a network of telescopes dedicated to high-cadence, high-precision ground-based RV observations of solar-like oscillators (\citealt{Andersen2014, Grundahl2017}; Lund et al., in prep.). Currently, two SONG nodes are active. One is the 1-m Danish Hertzsprung SONG Telescope at Observatorio del Teide in Tenerife \citep[SONG-Tenerife;][]{Grundahl2017}, and the other is at the Mount Kent Observatory near Toowoomba, Queensland, Australia \citep[SONG-Australia;][]{Kjeldsen2025}. Furthermore, while SONG's primary goal is to obtain high-cadence observations of solar-like oscillators  \citep[e.g.][]{Grundahl2017, Malla2020, Knudstrup2023, Kjeldsen2025}, it has also been used for long-term RV monitoring of other intrinsically variable stars \citep[e.g.][]{Spaeth2025}, binaries (e.g. \citealt{Brogaard2021}; Rudrasingam et al., in prep.), and exoplanet host stars \citep[e.g.][]{Luque2019, Addison2021}. 

We identified 69 stars that have observations from SONG-Tenerife, of which 29 have more than 60 spectra. In \autoref{tab:song_list}, we list the number of stars in our sample with SONG observations up to and including 2024. \new{We also list 19 stars in \autoref{tab:song_overlap_list} that have simultaneous TESS and SONG observations. These observations can be used for both forward modelling of stellar variability with simultaneous photometry and RV, thereby removing the stellar signal from RV observations for exoplanet detection \citep[e.g.][]{Beard2025, Tang2026}. For solar-like oscillators, some of these observations can be used to measure the amplitude ratio and phase difference between photometry and RV \citep[e.g.][]{Jimenez2002, Huber2011, Kjeldsen2025}.} The SONG data themselves are hosted on SODA\footnote{\url{https://soda.phys.au.dk/}}, and are freely available upon registration.

\section{Conclusions}\label{sec:conclusion}

We extracted light curves of 98 of the brightest stars observed by TESS in Sectors 1--93 using halo photometry \citep{White2017, Pope2019}. We used an updated version of the \textsc{halophot}, written in the \textsc{Python} framework \textsc{Jax} \citep{jax2018} framework, to extract the light curves from a $31\times31$ cut of the Full Frame Images (FFIs) for each star. 

We compared the quality of the \textsc{halophot} light curves with Science Processing Operations Center \citep[SPOC;][]{Jenkins2016}, TESS-SPOC \citep{Caldwell2020}, Quick-look Pipeline \citep[QLP;][]{Huang2020a, Huang2020b, Kunimoto2021, Kunimoto2022}, and non-Total Variation minimised light curves, and found that the \textsc{halophot} light curves have comparable quality compared to the SPOC light curves, but that the \textsc{halophot} light curves are of higher quality than SPOC at the bright end of our sample. In contrast, for fainter stars ($V > 2.55$), we observe the opposite trend. A sizeable number of light curves have comparable quality between \textsc{halophot} and SPOC, but the SPOC light curves had underestimated amplitudes due to aperture masks that included saturated pixels. The QLP light curves are mostly worse than the \textsc{halophot} light curves.

We detected variability in 15 red giants, and we derived the global asteroseismic parameter, \numax{}, using \textsc{pyMON} \citep{Howell2025} for 13 of them. We then used the derived \numax{} and literature radii to estimate their masses. Among the 13 red giants, five of them are newly discovered solar-like oscillators. For one red giant, $\beta$~Gem (Pollux), we measured \Dnu{} and derived its seismic mass and radius.

Among the A and early F stars, we detected one \gdor{} variable and five \dsct{} variables, of which two are newly discovered. One of them, $\alpha$~Cep (Alderamin), is a newly discovered \dsct{}--\gdor{} hybrid pulsator. For the O and B stars, we detected nine $\beta$~Cephei variables, three slowly pulsating B-type stars, eight Be stars, and eight stochastic low-frequency variables. Among them is the binary system $\epsilon$~Car (Avior), which we found to be system consisting of a long-period variable and either a slowly pulsating B-type star or a $\alpha^2$~CVn variable. We also detected eight eclipsing binaries, with one of them, $\gamma$~And (Almach), being a newly discovered eclipsing binary. 

We compared the \textsc{halophot} results from TESS and $K$2 for stars in common \citep{Pope2019} and found good agreement. We found agreement between the depths of the primary eclipses for three of our EBs with published values.

We identified nine stars that will be inside the PLAnetary Transits and Oscillations of stars \citep[PLATO;][]{Rauer2025} southern Long-duration Observation Phase (LOP) and four that will be inside its northern LOP. Two stars, $\gamma$~Dra and $\alpha$~Col, will be within the field of view of the fast cameras, potentially enabling multi-colour photometry for both.

Finally, we identified 69 stars in our sample with spectra from the Stellar Observations Network Group \citep[SONG;][]{Andersen2014, Grundahl2017, Kjeldsen2025}, of which 29 have more than 60 spectra. Furthermore, 19 stars have simultaneous SONG and TESS observations, enabling forward modelling of stellar variability and, for a few of them, measuring the amplitude ratio and phase difference between photometry and RV.

Altogether, we find 77 out of 98 stars to be variable and provide high-quality \textsc{halophot} light curves on MAST for easy access and future detailed analysis.

\section*{Acknowledgements}

We are grateful for discussions with Courtney Crawford about the stochastic nature of $\theta$~Sco, Jeremy Bailey about stochastic variability in blue supergiants, and Ben Montet about TESS photometry.

We acknowledge and pay respect to the traditional owners of the land on which the University of Sydney and Macquarie University are situated, the Gadigal clan of the Eora Nation and the Wallumattagal clan of the Dharug Nation, upon whose unceded, sovereign, ancestral lands we work. We pay respects to their Ancestors and descendants, who continue cultural and spiritual connections to Country. 

JR and TRB have been supported by the Australian Research Council through the Laureate Fellowship FL220100117, BP by the grants DP230101439 and DE210101639, and MGP by the Discovery Early Career Researcher Award (project number DE250100146). We are grateful to the Australian public for enabling this science. MNL acknowledges support from the ESA PRODEX programme (PEA 4000142995).

This paper includes data collected by the TESS mission \citep{Ricker2015} and $K$2 mission \citep{Howell2014}. Funding for the TESS and $K$2 missions is provided by the NASA's Science Mission Directorate. This paper made use of the Hipparcos and Tycho catalogues. Funding for the Hipparcos mission was provided by the European Space Agency.  This paper has made use of the SIMBAD\footnote{\url{https://simbad.u-strasbg.fr/simbad/sim-basicIdent=m33&submit=SIMBAD+search}} database \citep{Wenger2000}, operated at CDS, Strasbourg, France. This paper has also made use of the SONG database SODA, operated and maintained at Aarhus University, DK.

This research made use of \textsc{NumPy} \citep{numpy2020}; \textsc{Matplotlib} \citep{Matplotlib2007}; \textsc{SciPy} \citep{SciPy2020}; \textsc{Astropy} \citep{Astropy1, Astropy2, Astropy3}; \textsc{Astroquery} \citep{astroquery}; \textsc{lightkurve} \citep{lightkurve2018}; \textsc{pandas} \citep{pandas}; \textsc{pyMON} \citep{Howell2025}; \textsc{nifty-ls} \citep{NIFTY2024}; \textsc{uncertainties} \citep{uncertainties}; \textsc{JAX} \citep{jax2018}; and \textsc{JAXopt} \citep{jaxopt_implicit_diff}.

\section*{Data Availability}

All codes and data used to produce this work are publicly available and open source. The TESS \textsc{halophot} module is hosted at \href{https://github.com/hvidy/halophot}{https://github.com/hvidy/halophot} and diagnostic plots are at \href{https://github.com/JonatanRudrasingam/TESS_halophot}{https://github.com/JonatanRudrasingam/TESS\_halophot}. All light curves are available from MAST via \url{https://doi.org/10.17909/tzm7-r151}\footnote{\url{https://archive.stsci.edu/hlsp/halo-tess}} and accessible under the name ``HALO-TESS''. 



\bibliographystyle{mnras}
\bibliography{reference} 




\appendix
\onecolumn
\newpage
\section{\new{List of stars}}\label{appendix:stars_list}
\begin{longtable}{llrrlrrlllll}
\caption{All TESS stars with Halo Photometry.}
\label{tab:stars_list}
\endfirsthead
\endhead
\hline
Star & Name & TIC & HD & $V$ & $B - V$ & $M_V$ & Spectral type & $N$ & $N_H$ & $N_P$ & Variability \\
\hline
$\alpha$ Eri & Achernar & 230981971 & 10144 & 0.45 & -0.16 & -2.77 & B3Vpe & 3 & 3 & 1 & Be \\
$\alpha$ Ori & Betelgeuse & 269273552 & 39801 & 0.45 & 1.50 & -5.14 & M1-2Ia-Iab & 3 & 0 & 0 & \\
$\beta$ Cen & Hadar & 328329822 & 122451 & 0.61 & -0.23 & -5.42 & B1III & 3 & 3 & 1 & $\beta$ Cep \\
$\alpha$ Aql & Altair & 70257116 & 187642 & 0.76 & 0.22 & 2.20 & A7V & 2 & 2 & 0 & $\delta$ Sct \\
$\alpha$ Cru & Acrux & 450568754 & 108248 & 0.77 & -0.24 & -4.19 & B0.5IV+B1V & 4 & 4 & 0 & $\beta$ Cep \\
$\alpha$ Tau & Aldebaran & 245873777 & 29139 & 0.87 & 1.54 & -0.63 & K5+III & 4 & 4 & 0 & RG \\
$\alpha$ Vir & Spica & 178999156 & 116658 & 0.98 & -0.23 & -3.55 & B1III-IV+B2V & 1 & 1 & 0 & $\beta$ Cep + ELL \\
$\alpha$ Sco & Antares & 175934060 & 148478 & 1.06 & 1.87 & -5.28 & M1.5Iab-Ib+B4Ve & 1 & 0 & 0 & \\
$\beta$ Gem & Pollux & 423088367 & 62509 & 1.16 & 0.99 & 1.09 & K0IIIb & 5 & 5 & 0 & RG \\
$\alpha$ PsA & Fomalhaut & 47552789 & 216956 & 1.17 & 0.14 & 1.74 & A3V & 3 & 0 & 0 &  \\
$\alpha$ Cyg & Deneb & 195554360 & 197345 & 1.25 & 0.09 & -8.73 & A2Ia & 6 & 6 & 0 & SLF \\
$\beta$ Cru & Mimosa & 405567821 & 111123 & 1.25 & -0.24 & -3.92 & B0.5III & 5 & 4 & 5 & $\beta$ Cep \\
$\alpha$ Leo & Regulus & 357348164 & 87901 & 1.36 & -0.09 & -0.52 & B7V & 1 & 0 & 0 &  \\
$\epsilon$ CMa & Adhara & 63198307 & 52089 & 1.50 & -0.21 & -4.10 & B2II & 5 & 5 & 5 & SLF \\
$\alpha$ Gem & Castor & 239187696 & 60179 & 1.58 & 0.03 & 0.59 & A2Vm+A1V & 8 & 6 & 8 & EB \\
$\gamma$ Cru & Gacrux & 272314138 & 108903 & 1.59 & 1.60 & -0.56 & M3.5III & 4 & 4 & 0 & LPV \\
$\lambda$ Sco & Shaula & 465088681 & 158926 & 1.62 & -0.23 & -5.05 & B2IV+B & 4 & 2 & 3 & $\beta$ Cep + EB \\
$\gamma$ Ori & Bellatrix & 365572007 & 35468 & 1.64 & -0.22 & -2.72 & B2III & 2 & 2 & 0 & Var \\
$\beta$ Tau & Elnath & 285473140 & 35497 & 1.65 & -0.13 & -1.37 & B7III & 4 & 0 & 4 & $\alpha^2$~CVn \\
$\beta$ Car & Miaplacidus & 290374453 & 80007 & 1.67 & 0.07 & -0.99 & A2IV & 10 & 0 & 5 & Var \\
$\epsilon$ Ori & Alnilam & 427451176 & 37128 & 1.69 & -0.18 & -6.38 & B0Ia & 2 & 2 & 2 & SLF \\
$\alpha$ Gru & Alnair & 279316667 & 209952 & 1.73 & -0.07 & -0.73 & B7IV & 2 & 0 & 0 &  \\
$\zeta$ Ori & Alnitak & 11360636 & 37742 & 1.74 & -0.20 & -5.26 & O9.7Ib+B0III & 1 & 1 & 1 & SLF \\
$\gamma^2$ Vel &  & 354825513 & 68273 & 1.75 & -0.14 & -5.31 & WC8+O9I & 8 & 7 & 8 & SLF \\
$\epsilon$ UMa & Alioth & 150387644 & 112185 & 1.76 & -0.02 & -0.21 & A0pCr & 6 & 6 & 0 & $\alpha^2$ CVn \\
$\epsilon$ Sgr & Kaus Australis & 66389641 & 169022 & 1.79 & -0.03 & -1.44 & B9.5III & 5 & 4 & 1 & Rot \\
$\alpha$ Per & Mirfak & 252830952 & 20902 & 1.79 & 0.48 & -4.50 & F5Ib & 3 & 0 & 0 &  \\
$\alpha$ UMa & Dubhe & 99843265 & 95689 & 1.81 & 1.06 & -1.08 & K0IIIa & 4 & 4 & 0 & RG \\
$\delta$ CMa & Wezen & 64602863 & 54605 & 1.83 & 0.67 & -6.87 & F8Ia & 5 & 0 & 0 &  \\
$\eta$ UMa & Alkaid & 219033887 & 120315 & 1.85 & -0.10 & -0.60 & B3V & 5 & 5 & 5 & Be \\
$\theta$ Sco & Sargas & 17158018 & 159532 & 1.86 & 0.41 & -2.75 & F1II & 2 & 2 & 0 & SLF? \\
$\epsilon$ Car & Avior & 342884451 & 71129 & 1.86 & 1.20 & -4.58 & K3III+B2:V & 12 & 12 & 0 & LPV + SPB/$\alpha^2$ CVn \\
$\beta$ Aur & Menkalinan & 440388263 & 40183 & 1.90 & 0.08 & -0.10 & A2IV & 3 & 3 & 0 & EB \\
$\alpha$ TrA & Atria & 364216056 & 150798 & 1.91 & 1.45 & -3.62 & K2IIb-IIIa & 3 & 3 & 1 & LPV \\
$\gamma$ Gem & Alhena & 308056612 & 47105 & 1.93 & 0.00 & -0.60 & A0IV & 7 & 0 & 0 &  \\
$\delta$ Vel & Alsephina & 45696212 & 74956 & 1.93 & 0.04 & -0.01 & A1V & 5 & 4 & 0 & EB \\
$\alpha$ Pav & Peacock & 219974785 & 193924 & 1.94 & -0.12 & -1.81 & B2IV & 3 & 3 & 0 & SPB \\
$\alpha$ UMi & Polaris & 303256075 & 8890 & 1.97 & 0.64 & -3.64 & F7:Ib-II & 14 & 14 & 3 & Cepheid \\
$\beta$ CMa & Mirzam & 34590771 & 44743 & 1.98 & -0.24 & -3.95 & B1II-III & 3 & 3 & 1 & $\beta$ Cep \\
$\alpha$ Hya & Alphard & 46799297 & 81797 & 1.99 & 1.44 & -1.69 & K3II-III & 4 & 4 & 0 & LPV \\
$\gamma$ Leo & Algieba & 95431294 & 89484 & 2.01 & 1.13 & -0.92 & G7IIIFe-1+K1-IIIbFe-0.5 & 4 & 4 & 0 & RG \\
$\alpha$ Ari & Hamal & 306349516 & 12929 & 2.01 & 1.15 & 0.48 & K2-IIICa-1 & 6 & 5 & 0 & RG \\
$\beta$ Cet & Diphda & 114434141 & 4128 & 2.04 & 1.02 & -0.30 & G9.5IIICH-1 & 2 & 0 & 0 &  \\
$\sigma$ Sgr & Nunki & 91093307 & 175191 & 2.05 & -0.13 & -2.14 & B2.5V & 1 & 1 & 1 & SPB \\
$\theta$ Cen & Menkent & 179323446 & 123139 & 2.06 & 1.01 & 0.70 & K0-IIIb & 3 & 3 & 0 & RG \\
$\beta$ And & Mirach & 174500619 & 6860 & 2.07 & 1.58 & -1.86 & M0+IIIa & 3 & 3 & 0 & LPV \\
$\beta$ UMi & Kochab & 229540730 & 131873 & 2.07 & 1.47 & -0.87 & K4-III & 16 & 16 & 0 & RG \\
$\kappa$ Ori & Saiph & 66651575 & 38771 & 2.07 & -0.17 & -4.65 & B0.5I & 2 & 2 & 0 & SLF \\
$\alpha$ And & Alpheratz & 427733653 & 358 & 2.07 & -0.04 & -0.30 & B8IVpMnHg & 3 & 3 & 2 & $\alpha^2$ CVn \\
$\beta$ Gru & Tiaki & 44577667 & 214952 & 2.07 & 1.61 & -1.52 & M5III & 3 & 3 & 0 & LPV \\
$\alpha$ Oph & Rasalhague & 289643770 & 159561 & 2.08 & 0.16 & 1.30 & A5III & 1 & 1 & 0 & $\delta$ Sct + $\gamma$ Dor \\
$\beta$ Per & Algol & 346783960 & 193565 & 2.09 & -0.00 & -0.18 & B8V & 3 & 3 & 0 & EB \\
$\gamma$ And & Almach & 292057658 & 12533 & 2.10 & 1.37 & -3.08 & K3-IIb & 3 & 3 & 0 & EB \\
$\beta$ Leo & Denebola & 14725877 & 102647 & 2.14 & 0.09 & 1.92 & A3V & 3 & 0 & 0 &  \\
$\gamma$ Cas &  & 51962733 & 5394 & 2.15 & -0.05 & -4.22 & B0IVe & 6 & 6 & 6 & Be \\
$\gamma$ Cen & Muhlifain & 161739042 & 110304 & 2.20 & -0.02 & -0.81 & A1IV & 4 & 4 & 4 & Rot \\
$\zeta$ Pup & Naos & 133422778 & 66811 & 2.21 & -0.27 & -5.95 & O5f & 8 & 7 & 8 & SLF \\
$\iota$ Car & Aspidiske & 386296645 & 80404 & 2.21 & 0.19 & -4.42 & A8Ib & 9 & 0 & 0 &  \\
$\alpha$ CrB & Alphecca & 274945059 & 139006 & 2.22 & 0.03 & 0.42 & A0V+G5V & 3 & 3 & 0 & EB \\
$\gamma$ Cyg & Sadr & 13431346 & 194093 & 2.23 & 0.67 & -6.12 & F8Ib & 7 & 0 & 0 &  \\
$\zeta$ UMa & Mizar & 159190005 & 116656 & 2.23 & 0.06 & 0.33 & A1m + A1VpSrSi & 5 & 0 & 0 &  \\
$\lambda$ Vel & Suhail & 31975064 & 78647 & 2.23 & 1.66 & -3.99 & K4.5Ib-II & 6 & 6 & 0 & LPV \\
$\gamma$ Dra & Eltanin & 329269366 & 164058 & 2.24 & 1.52 & -1.04 & K5III & 18 & 18 & 0 & RG \\
$\alpha$ Cas & Schedar & 312141546 & 3712 & 2.24 & 1.17 & -1.99 & K0IIIa & 5 & 5 & 1 & RG \\
$\delta$ Ori & Mintaka & 50743469 & 36485 & 2.25 & -0.17 & -4.99 & O9.5II & 2 & 2 & 0 & EB \\
$\beta$ Cas & Caph & 396298498 & 432 & 2.28 & 0.38 & 1.17 & F2III-IV & 9 & 9 & 8 & $\delta$ Sct \\
$\epsilon$ Cen &  & 241398115 & 118716 & 2.29 & -0.17 & -3.02 & B1III & 3 & 3 & 0 & $\beta$ Cep \\
$\epsilon$ Sco & Larawag & 191437754 & 151680 & 2.29 & 1.14 & 0.78 & K2.5III & 3 & 3 & 0 & RG \\
$\delta$ Sco & Dschubba & 12725034 & 143275 & 2.29 & -0.12 & -3.16 & B0.3IV & 1 & 1 & 1 & Be \\
$\alpha$ Lup & Uridim & 129117325 & 129056 & 2.30 & -0.15 & -3.83 & B1.5III/Vn & 3 & 3 & 1 & $\beta$ Cep \\
$\eta$ Cen &  & 128116539 & 127972 & 2.33 & -0.16 & -2.55 & B1.5Vne & 3 & 2 & 2 & Be \\
$\beta$ UMa & Merak & 141120277 & 95418 & 2.34 & 0.03 & 0.41 & A1V & 3 & 0 & 0 &  \\
$\epsilon$ Boo & Izar & 219827143 & 129988 & 2.35 & 0.97 & -1.69 & K0-II-III & 3 & 3 & 0 & RG \\
$\epsilon$ Peg & Enif & 466337046 & 206778 & 2.38 & 1.52 & -4.19 & K2Ib & 2 & 0 & 0 & \\
$\kappa$ Sco & Girtab & 147868882 & 160578 & 2.39 & -0.17 & -3.38 & B1.5III & 3 & 0 & 3 & $\beta$ Cep \\
$\alpha$ Phe & Ankaa & 80256524 & 2261 & 2.40 & 1.08 & 0.52 & K0III & 2 & 2 & 1 & RG \\
$\gamma$ UMa & Phecda & 11895653 & 103287 & 2.41 & 0.04 & 0.36 & A0Ve & 5 & 0 & 0 &  \\
$\eta$ Oph & Sabik & 400036304 & 155125 & 2.43 & 0.06 & 0.37 & A2V & 1 & 0 & 0 &  \\
$\beta$ Peg & Scheat & 436774002 & 217906 & 2.44 & 1.65 & -1.49 & M2.5II-III & 2 & 2 & 0 & LPV \\
$\alpha$ Cep & Alderamin & 417604820 & 203280 & 2.45 & 0.26 & 1.58 & A7V & 11 & 11 & 2 & $\delta$ Sct + $\gamma$ Dor \\
$\eta$ CMa & Aludra & 107415639 & 58350 & 2.45 & -0.08 & -7.51 & B5Ia & 4 & 4 & 0 & SLF \\
$\kappa$ Vel & Markeb & 387106852 & 81188 & 2.47 & -0.14 & -3.62 & B2IV-V & 6 & 6 & 0 & SPB \\
$\epsilon$ Cyg & Aljanah & 232853959 & 197989 & 2.48 & 1.02 & 0.76 & K0-III & 5 & 5 & 0 & RG \\
$\zeta$ Oph &  & 152859121 & 149757 & 2.54 & 0.04 & -3.20 & O9.5Vn & 1 & 1 & 0 & Be \\
$\zeta$ Cen & Leepwal & 113350416 & 121263 & 2.55 & -0.18 & -2.81 & B2.5IV & 3 & 3 & 0 & Heartbeat star \\
$\delta$ Leo & Zosma & 159523958 & 97603 & 2.56 & 0.13 & 1.32 & A4V & 2 & 2 & 2 & Rot \\
$\gamma$ Crv & Gienah & 348987372 & 106625 & 2.58 & -0.11 & -0.94 & B8IIIpHgMn & 4 & 1 & 3 & Rot \\
$\delta$ Cen &  & 333670784 & 105435 & 2.58 & -0.13 & -2.84 & B2IVne & 3 & 2 & 3 & Be \\
$\alpha$ Lep & Arneb & 46312112 & 36673 & 2.58 & 0.21 & -5.40 & F0Ib & 1 & 0 & 0 &  \\
$\zeta$ Sgr & Ascella & 30504485 & 176687 & 2.60 & 0.06 & 0.42 & A2III+A4IV & 3 & 0 & 0 &  \\
$\beta$ Lib & Zubeneschamali & 79391147 & 135742 & 2.61 & -0.07 & -0.84 & B8V & 2 & 0 & 0 &  \\
$\alpha$ Ser & Unukalhai & 296856546 & 140573 & 2.63 & 1.17 & 0.87 & K2IIIbCN1 & 1 & 1 & 1 & RG \\
$\beta$ Ari & Sheratan & 91356756 & 11636 & 2.64 & 0.17 & 1.33 & A5V & 5 & 0 & 5 & $\gamma$ Dor \\
$\beta$ Crv & Kraz & 83151778 & 109379 & 2.65 & 0.89 & -0.51 & G5II & 1 & 0 & 0 &  \\
$\alpha$ Col & Phact & 140214221 & 37795 & 2.65 & -0.12 & -1.93 & B7IVe & 5 & 0 & 5 & Be \\
$\theta$ Aur & Mahasim & 174671483 & 40312 & 2.65 & -0.08 & -0.98 & A0pSi & 2 & 2 & 1 & $\alpha^2$ CVn \\
$\delta$ Cas & Ruchbah & 54995745 & 8538 & 2.66 & 0.16 & 0.24 & A5III-IV & 3 & 1 & 3 & $\delta$ Sct \\
$\beta$ Her & Kornephoros & 284853659 & 148856 & 2.78 & 0.95 & -0.50 & G7IIIa & 5 & 5 & 0 & RG \\
\hline
\end{longtable}
\newpage
\section{Stars with SONG observations}\label{appendix:song_list}
Here we list the stars with SONG observations. In the first table (\autoref{tab:song_list}), we list the star, HIP number, apparent magnitude in $V$, the number of spectra, and our TESS variability classification. In the second table (\autoref{tab:song_overlap_list}), we list the stars that have simultaneous SONG and TESS observations. Here we show the star, HIP number, number of simultaneous observations, median time difference in minutes, the Nyquist frequency in \muhz{}, and our TESS variability classification. The SONG observations were obtained from SODA, and includes all observations up to and including 2024.
\begin{longtable}{lrlll}
\caption{Stars with SONG observation}
\label{tab:song_list}
\endfirsthead
\endhead
\hline
Star & HIP & $V$ & $N_{\rm spectra}$ & Variability \\
\hline
$\theta$ Cen & 68933 & 2.06 & 7669 & RG \\
$\beta$ Gem & 37826 & 1.16 & 7609 & RG \\
$\epsilon$ Ori & 26311 & 1.69 & 5583 & SLF \\
$\epsilon$ Cyg & 102488 & 2.48 & 5359 & RG \\
$\lambda$ Sco & 85927 & 1.62 & 4296 & $\beta$ Cep + EB \\
$\alpha$ Tau & 21421 & 0.87 & 3234 & RG \\
$\alpha$ Ori & 27989 & 0.45 & 1331 & LPV \\
$\beta$ Per & 14576 & 2.09 & 1237 & EB \\
$\kappa$ Ori & 27366 & 2.07 & 1196 & SLF \\
$\delta$ Ori & 25930 & 2.25 & 1094 & EB \\
$\beta$ Cas & 746 & 2.28 & 770 & $\delta$ Sct \\
$\alpha$ Ari & 9884 & 2.01 & 761 & RG \\
$\theta$ Aur & 28380 & 2.65 & 707 & $\alpha^2$ CVn \\
$\alpha$ UMi & 11767 & 1.97 & 634 & Cepheid \\
$\beta$ Aur & 28360 & 1.90 & 619 & EB \\
$\gamma$ Dra & 87833 & 2.24 & 590 & RG \\
$\gamma$ Leo & 50583 & 2.01 & 574 & RG \\
$\alpha$ Sco & 80763 & 1.06 & 545 & LPV \\
$\beta$ UMi & 72607 & 2.07 & 486 & RG \\
$\epsilon$ Boo & 72105 & 2.35 & 463 & RG \\
$\alpha$ UMa & 54061 & 1.81 & 428 & RG \\
$\alpha$ Leo & 49669 & 1.36 & 365 &  \\
$\alpha$ CrB & 76267 & 2.22 & 323 & EB \\
$\alpha$ Hya & 46390 & 1.99 & 307 & LPV \\
$\zeta$ UMa & 65378 & 2.23 & 283 &  \\
$\gamma$ Cas & 4427 & 2.15 & 233 & Be \\
$\eta$ UMa & 67301 & 1.85 & 203 & Be \\
$\alpha$ And & 677 & 2.07 & 88 & $\alpha^2$ CVn \\
$\gamma$ Cyg & 100453 & 2.23 & 61 &  \\
$\alpha$ Cas & 3179 & 2.24 & 6 & RG \\
$\zeta$ Oph & 81377 & 2.54 & 6 & Be \\
$\beta$ Leo & 57632 & 2.14 & 6 &  \\
$\delta$ Leo & 54872 & 2.56 & 6 & Rot \\
$\alpha$ Oph & 86032 & 2.08 & 6 & $\delta$ Sct + $\gamma$ Dor \\
$\alpha$ Cyg & 102098 & 1.25 & 5 & SLF \\
$\gamma$ Gem & 31681 & 1.93 & 4 &  \\
$\beta$ And & 5447 & 2.07 & 4 & LPV \\
$\eta$ CMa & 35904 & 2.45 & 4 & SLF \\
$\delta$ Cas & 6686 & 2.66 & 3 & $\delta$ Sct \\
$\gamma$ Crv & 59803 & 2.58 & 3 & Rot \\
$\alpha$ Aql & 97649 & 0.76 & 3 & $\delta$ Sct \\
$\beta$ Crv & 61359 & 2.65 & 3 &  \\
$\alpha$ Vir & 65474 & 0.98 & 3 & $\beta$ Cep + ELL \\
$\beta$ Ari & 8903 & 2.64 & 3 & $\gamma$ Dor \\
$\alpha$ Ser & 77070 & 2.63 & 3 & RG \\
$\alpha$ PsA & 113368 & 1.17 & 3 &  \\
$\beta$ Lib & 74785 & 2.61 & 3 &  \\
$\zeta$ Sgr & 93506 & 2.60 & 3 &  \\
$\epsilon$ CMa & 33579 & 1.50 & 3 & SLF \\
$\alpha$ Gem & 36850 & 1.58 & 3 & EB \\
$\alpha$ Lep & 25985 & 2.58 & 3 &  \\
$\gamma$ Ori & 25336 & 1.64 & 3 & Var \\
$\beta$ Tau & 25428 & 1.65 & 3 & Rot \\
$\zeta$ Ori & 26727 & 1.74 & 3 & SLF \\
$\gamma$ And & 9640 & 2.10 & 3 & EB \\
$\epsilon$ UMa & 62956 & 1.76 & 3 & $\alpha^2$ CVn \\
$\alpha$ Per & 15863 & 1.79 & 3 &  \\
$\delta$ CMa & 34444 & 1.83 & 3 &  \\
$\alpha$ Cep & 105199 & 2.45 & 3 & $\delta$ Sct + $\gamma$ Dor \\
$\beta$ Peg & 113881 & 2.44 & 3 & LPV \\
$\eta$ Oph & 84012 & 2.43 & 3 &  \\
$\gamma$ UMa & 58001 & 2.41 & 3 &  \\
$\beta$ CMa & 30324 & 1.98 & 3 & $\beta$ Cep \\
$\epsilon$ Peg & 107315 & 2.38 & 3 & LPV \\
$\beta$ UMa & 53910 & 2.34 & 3 &  \\
$\delta$ Sco & 78401 & 2.29 & 3 & Be \\
$\beta$ Cet & 3419 & 2.04 & 3 &  \\
$\sigma$ Sgr & 92855 & 2.05 & 3 & SPB \\
$\beta$ Her & 80816 & 2.78 & 3 & RG \\
\hline
\end{longtable}
\begin{longtable}{lrllll}
\caption{Simultaneous TESS and SONG stars}
\label{tab:song_overlap_list}
\endfirsthead
\endhead
\hline
Star & HIP & Simultaneous & $\Delta t \ (\rm min) $ & $\nu_{\rm Nyquist}$ ($\mu \rm Hz$) & Variability \\
\hline
$\theta$ Cen & 68933 & 5419 & 1.088 & 7659.314 & RG \\
$\beta$ Gem & 37826 & 68 & 0.812 & 10262.726 & RG \\
$\epsilon$ Cyg & 102488 & 2 & 1031980.025 & 0.008 & RG \\
$\lambda$ Sco & 85927 & 1181 & 6.088 & 1368.813 & $\beta$ Cep + EB \\
$\alpha$ Tau & 21421 & 153 & 0.68 & 12254.902 & RG \\
$\beta$ Per & 14576 & 7 & 1903.868 & 4.377 & EB \\
$\alpha$ Ari & 9884 & 98 & 1.845 & 4516.712 & RG \\
$\alpha$ UMi & 11767 & 324 & 2.114 & 3941.974 & Cepheid \\
$\beta$ Aur & 28360 & 58 & 29.81 & 279.548 & EB \\
$\gamma$ Dra & 87833 & 112 & 1755.446 & 4.747 & RG \\
$\gamma$ Leo & 50583 & 23 & 2917.931 & 2.856 & RG \\
$\beta$ UMi & 72607 & 198 & 0.63 & 13227.513 & RG \\
$\epsilon$ Boo & 72105 & 48 & 1413.474 & 5.896 & RG \\
$\alpha$ UMa & 54061 & 3 & 1436.742 & 5.8 & RG \\
$\alpha$ CrB & 76267 & 18 & 85.955 & 96.95 & EB \\
$\alpha$ Hya & 46390 & 36 & 258.096 & 32.288 & LPV \\
$\zeta$ UMa & 65378 & 27 & 1479.122 & 5.634 &  \\
$\eta$ UMa & 67301 & 29 & 1.09 & 7645.26 & Be \\
$\gamma$ Cyg & 100453 & 8 & 7390.715 & 1.128 &  \\
\hline
\end{longtable}


\bsp	
\label{lastpage}
\end{document}